\documentclass[sigconf, nonacm]{acmart}

\usepackage{enumitem}
\usepackage{algorithm}
\usepackage[noend]{algpseudocode}
\usepackage{multirow}
\usepackage{subfig}
\usepackage{booktabs}   
\usepackage{pifont}     
\usepackage{array}

\newcommand{\cmark}{\ding{51}}   
\newcommand{\xmark}{\ding{55}}   

\newcommand\vldbdoi{XX.XX/XXX.XX}
\newcommand\vldbpages{XXX-XXX}
\newcommand\vldbvolume{14}
\newcommand\vldbissue{1}
\newcommand\vldbyear{2020}
\newcommand\vldbauthors{\authors}
\newcommand\vldbtitle{\shorttitle} 
\newcommand\vldbavailabilityurl{URL_TO_YOUR_ARTIFACTS}
\newcommand\vldbpagestyle{plain} 

\begin{document}
\title{Filtered Approximate Nearest Neighbor Search: A Unified Benchmark and Systematic Experimental Study [Experiment, Analysis \& Benchmark]}
\author{Jiayang Shi}
\affiliation{%
  \institution{Fudan University, Shanghai, China}
}
\email{jiayangshi22@m.fudan.edu.cn}

\author{Yuzheng Cai}
\affiliation{%
  \institution{Fudan University, Shanghai, China}
}
\email{yuzhengcai21@m.fudan.edu.cn}

\author{Weiguo Zheng}
\affiliation{%
  \institution{Fudan University, Shanghai, China}
}
\email{zhengweiguo@fudan.edu.cn}





\begin{abstract}
  For a given dataset $\mathcal{D}$ and structured label $f$, the goal of Filtered Approximate Nearest Neighbor Search (FANNS) algorithms is to find top-$k$ points closest to a query that satisfy label constraints, while ensuring both recall and QPS (Queries Per Second). In recent years, many FANNS algorithms have been proposed. However, the lack of a systematic investigation makes it difficult to understand their relative strengths and weaknesses. Additionally, we found that: (1) FANNS algorithms have coupled, dataset-dependent parameters, leading to biased comparisons. (2) Key impact factors are rarely analyzed systematically, leaving unclear when each algorithm performs well. (3) Disparate datasets, workloads, and biased experiment designs make cross-algorithm comparisons unreliable.
  

    Thus, a comprehensive survey and benchmark for FANNS is crucial to achieve the following goals: designing a fair evaluation and clarifying the classification of algorithms, conducting in-depth analysis of their performance, and establishing a unified benchmark. First, we propose a taxonomy (dividing methods into \textit{filter-then-search}, \textit{search-then-filter}, \textit{hybrid-search}) and a systematic evaluation framework, integrating unified parameter tuning and standardized filtering across algorithms to reduce implementation-induced performance variations and reflect core trade-offs. Then, we conduct a comprehensive empirical study to analyze how query difficulty and dataset properties impact performance, evaluating robustness under pressures like filter selectivity, Recall@k, and scalability to clarify each method’s strengths. Finally, we establish a standardized benchmark with real-world datasets and open-source related resources to ensure reproducible future research.

\end{abstract}

\maketitle

\pagestyle{\vldbpagestyle}
\begingroup
\vspace{-0.2cm}
\small\noindent\raggedright\textbf{PVLDB Reference Format:}\\
\vldbauthors. \vldbtitle. PVLDB, \vldbvolume(\vldbissue): \vldbpages, \vldbyear.\\
\href{https://doi.org/\vldbdoi}
\endgroup
\begingroup
\renewcommand\thefootnote{}\footnote{\noindent
This work is licensed under the Creative Commons BY-NC-ND 4.0 International License. Visit \url{https://creativecommons.org/licenses/by-nc-nd/4.0/} to view a copy of this license. For any use beyond those covered by this license, obtain permission by emailing \href{mailto:info@vldb.org}{info@vldb.org}. Copyright is held by the owner/author(s). Publication rights licensed to the VLDB Endowment. \\
\raggedright Proceedings of the VLDB Endowment, Vol. \vldbvolume, No. \vldbissue\ %
ISSN 2150-8097. \\
\href{https://doi.org/\vldbdoi}{doi:\vldbdoi} \\
}\addtocounter{footnote}{-1}\endgroup

\ifdefempty{\vldbavailabilityurl}{}{
\begingroup
\vspace{-0.3cm}
\small\noindent\raggedright\textbf{PVLDB Artifact Availability:}\\
The source code, data, and/or other artifacts have been made available at \url{https://github.com/sjyouuuuug/filterbenchmark}.
\endgroup
}

\section{Introduction}
\vspace{-.5mm}

Recent advances in recommendation systems~\cite{dssm, mind}, search engines~\cite{searchengine3, searchengine4}, and AI systems~\cite{RAGsurvey2023, RAGsurvey2024} have driven a growing need to retrieve and process large-scale, multi-modal unstructured data, including text, images, and video.
The most popular solution is to unify different modals into high-dimensional embeddings, then apply Approximate Nearest Neighbor Search (ANNS) algorithms to search semantically similar items in high-dimensional spaces~\cite{ANNSsurvey1, ANNSsurvey2, ANNSsurvey3, NSG, diskann}.
Furthermore, besides solely utilizing semantic similarity as the retrieval criteria, real-world applications \cite{recommandation1, recommandation2, recommandation3} also increasingly demand
%
hybrid query processing that integrates semantic search with metadata filtering, 
such as e-commerce systems~\cite{ecommerce1, ecommerce2} retrieving similar products within specified price, or academic search platforms identifying topically relevant publications filtered by publication date and venue.
This task combines vector similarity search with structured attribute constraints, and is commonly known as Filtered Approximate Nearest Neighbor Search (FANNS) \cite{UNG, filterDiskANN} or constrained Approximate Nearest Neighbor Search \cite{CAPS, HQANN}.

\vspace{-1.5mm}
\subsection{Background}
\vspace{-.5mm}

%
Filtered Approximate Nearest Neighbor Search (FANNS) is challenging due to the separated label spaces and vector spaces.
Naive implementations directly split label filtering and vector retrieval into distinct steps, which apply Boolean filters either pre- or post-ANNS search.
 However, such intuitive methods
suffer from fundamental efficiency limitations. Hybrid approaches remain suboptimal due to substantial computational overhead: systems frequently retrieve and process high-dimensional vectors that satisfy proximity requirements but fail attribute predicates, resulting in unnecessary distance computations and increased query latency~\cite{filterDiskANN, NHQ}.

Despite the proliferation of FANNS algorithms, a clear, comparative understanding of their practical performance remains elusive due to three fundamental challenges in the existing works.

\noindent \textit{\textbf{Combinatorial Explosion of Algorithmic Configurations.}}
The objective evaluation of algorithms is hindered by complex and coupled parameter spaces. As illustrated in Table~\ref{tab:algorithm_params_en_vline}, most methods expose a relatively large number of tuning parameters. These parameters are often strongly interdependent~\cite{vdtuner}, and their optimal configuration can vary dramatically with changes across datasets. 
This frequently leads to reported performance reflecting expert-level hyperparameter optimization while baseline methods may receive suboptimal configurations (with default or superficial parameters), obscuring true algorithmic merit. 
Establishing systematic evaluation protocols that decouple parameter tuning effects from intrinsic algorithmic advantages is therefore essential.



\noindent\textit{\textbf{Diverse Characteristics of Queries and Datasets.}}
The precise impact of query characteristics and dataset properties on FANNS performance remains insufficiently characterized.
And prior studies \cite{filterDiskANN, acorn, NHQ} frequently neglect systematic analysis of query difficulty despite performance being governed by multifaceted factors: filter selectivity, predicate complexity, target recall thresholds ($Recall@k$), and degradation patterns under dataset scaling \cite{UNG, CAPS}. 
A comprehensive investigation is needed to understand how different algorithms respond to these varying query- and data-specific pressures, 
identifying robust approaches across diverse scenarios.


\noindent\textit{\textbf{The Hidden Bias and Inconsistency in Evaluation.}}
The experimental setups, datasets, and filter workloads used in existing research~\cite{NHQ, filterDiskANN, CAPS} are highly fragmented, as different papers often employ varied datasets and experimental scenarios. This inconsistency is compounded by implicit design bias favoring newly proposed algorithms \cite{UNG, acorn}, manifested through non-standardized selectivity regimes, where some studies emphasize high-selectivity scenarios while others focus on low-selectivity cases or single-label constraints. The absence of unified benchmarking standards impedes reliable cross-algorithm comparison and obstructs definitive identification of the advantages of these techniques.



\begin{table}[t]
  \centering
  \caption{Dimension of Parameter Space for FANNS Methods}
  \vspace{-2.5mm}
  \label{tab:algorithm_params_en_vline}
  \setlength{\tabcolsep}{7pt}
  \begin{tabular}{lc|lc}
    \toprule
    \textbf{Algorithm} & \textbf{Dim} & \textbf{Algorithm} & \textbf{Dim} \\
    \midrule
    Pre-filter Brute Force & 0   & Post-filter HNSW & 2 \\
    Post-filter IVFPQ & 3   & ACORN-$\gamma$ & 3 \\
    ACORN-$1$ & 2   & UNG & 4 \\
    Filtered DiskANN &  3  & Stitched DiskANN & 4 \\
    NHQ-kgraph & 9 & CAPS & 2 \\
    \bottomrule
  \end{tabular}
  \vspace{-1mm}
\end{table}

\vspace{-1mm}
\subsection{Contributions}
\label{sec: contributions}
\vspace{-.5mm}

To address the challenges above, we conduct a comprehensive benchmark evaluation of FANNS approaches across multiple dimensions. The systematic assessment spans diverse datasets, filter selectivity regimes, and query workloads to characterize fundamental tradeoffs and operational constraints. 

   \noindent \textbf{A Framework for Fair, Parameter-Aware Evaluation.} 
    We introduce a systematic evaluation framework designed to ensure fairness and parameter-aware comparisons, thereby mitigating the evaluation bias often introduced by ad-hoc tuning. We establish an intuitive taxonomy that categorizes methods into three distinct classes: \textit{filter-then-search}, \textit{search-then-filter}, and \textit{hybrid-search}, which serves to elucidate their inherent algorithmic trade-offs. To enable a rigorous and objective comparison, we integrate a unified parameter tuning algorithm and a standardized filtering implementation, supporting both pre-computed bitsets and post-filtering strategies, across all evaluated methods. We evaluated $9$ algorithms across small subsets of $6$ datasets, testing over $41,000$ parameter combinations to build small indices and perform hybrid search. From these, we selected approximately $1,300$ representative parameter sets to construct full indexes on the entire datasets. This methodological rigor, which includes a comprehensive exploration of parameter combinations and index configurations, minimizes performance variance attributable to implementation-specific optimizations. Consequently, our benchmarking results isolate and accurately reflect the fundamental performance characteristics and core trade-offs of each algorithmic paradigm.

  \noindent \textbf{In-Depth Comparison Across Diverse Query and Dataset Characteristics.}
We conduct a comprehensive empirical study characterizing how query difficulty and dataset properties influence search performance. Our experimental design systematically evaluates algorithmic robustness across controlled operational pressures, including filter selectivity gradients, recall thresholds (Recall@k), and scalability limits. For example, we test performance under data scalability pressures by progressively increasing the dataset size from $5$ thousand to $5$ million items in Section~\ref{sec: base experiment}, which reveals the practical robustness and specific strengths of each approach. This rigorous assessment reveals comparative strengths and operational boundaries of each approach under realistic deployment conditions. 

    \noindent \textbf{A Unified and Reproducible Benchmark.}
%
To address the fragmentation and incomparability of existing experimental methodologies, we establish a standardized benchmark. This benchmark integrates multiple large-scale, real-world datasets with diverse characteristics and authentic labeling structures, such as YFCC\cite{YFCC} and  YouTube\cite{youtube}, and defines a comprehensive suite of filter workloads. Our open-source release of all datasets, evaluation scripts, and unified algorithm implementations provides a foundational resource for transparent and reproducible evaluation, ensuring that future research advances are built upon consistent and equitable experimental foundations.

In summary, we make the following contributions in this paper.
\begin{itemize}[leftmargin=*]
    \item 
    
To enable objective benchmarking, we present a novel evaluation framework that alleviates parameter tuning bias, thereby providing a clear analysis of inherent algorithmic trade-offs.
    
    \item We conduct a comprehensive experimental study to reveal the practical strengths and operational boundaries of different algorithms across diverse query and data scenarios.
    
    \item 
    We present and open-source a unified benchmark featuring standardized datasets and evaluation scripts, designed to foster transparency, reproducibility, and rigorous comparative analysis.

\end{itemize}

\section{Preliminary}
\subsection{Problem Definition}
\label{sec: problem definition}


We present the definitions of approximate nearest neighbor search and filtered approximate nearest neighbor search.

\begin{table*}[ht]
\centering
\caption{Comparison of support for different logical predicates across various algorithms. Abbreviations: CONT (Containment), OVER (Overlap), EQ (Equality), FIXED-EQ (Fixed Length Equality), COMB (Combination). COMB refers to the ability to handle complex logical expressions involving multiple predicates.}
\vspace{-2mm}
\label{tab:predicate_support}
 \setlength{\tabcolsep}{12pt}
\begin{tabular}{lcccccc}
\toprule
\textbf{METHOD} & \textbf{CONT} & \textbf{OVER} & \textbf{EQ} & \textbf{FIXED-EQ} & \textbf{COMB} & \textbf{Method Paradigm} \\  
\midrule
NHQ~\cite{NHQ} & \xmark & \xmark & \xmark & \cmark & \xmark & Hybrid-Search \\
FILTERED-DISKANN~\cite{filterDiskANN} & \cmark & \cmark & \cmark & \cmark & \cmark & Hybrid-Search \\
STITCHED-DISKANN~\cite{filterDiskANN} & \cmark & \cmark & \cmark & \cmark & \cmark & Hybrid-Search \\
ACORN-1~\cite{acorn} & \cmark & \cmark & \cmark & \cmark & \cmark & Filter-then-Search \\
ACORN-\(\gamma\)~\cite{acorn} & \cmark & \cmark & \cmark & \cmark & \cmark & Filter-then-Search \\
CAPS~\cite{CAPS} & \xmark & \xmark & \xmark & \cmark & \xmark & Hybrid-Search \\
UNG~\cite{UNG} & \cmark & \cmark & \cmark & \cmark & \xmark & Filter-then-Search \\
BRUTE-FORCE~\cite{UNG} & \cmark & \cmark & \cmark & \cmark & \cmark & Filter-then-Search \\
Post-filter HNSW~\cite{HNSW} & \cmark & \cmark & \cmark & \cmark & \cmark & Search-then-Filter \\
Post-filter IVFPQ~\cite{ivfpq} & \cmark & \cmark & \cmark & \cmark & \cmark & Search-then-Filter \\
\bottomrule
\end{tabular}
\end{table*}

\vspace{-1mm}
\begin{definition}[Approximate Nearest Neighbor Search, ANNS]
    Let \(\mathcal{D}\subset\mathbb{R}^d\) be a base dataset of \(n\) vectors and let \(v_q\in\mathbb{R}^d\) be a query vector.  Given an integer \(k\ge1\), the approximate \(k\)\nobreakdash-nearest neighbor search algorithm returns an approximate $k$-nearest neighbor set ($k$-NN) of size $k$,
\(
\mathcal{R}(v_q)\;=\;\{x_1,\dots,x_k\}\;\subseteq\;\mathcal{D},
\)
to minimize the distance of $\delta(v_q, v_i)$ for any $v_i\in \mathcal{R}(v_q)$ while reducing latency. 
\end{definition}
\vspace{-1mm}

We now extend it to define FANNS, where each vector $v_i \in \mathcal{D}$ 
is associated with a set of labels $f_i$. A filtered query must find vectors that are close in distance and satisfy a given label constraint~\cite{filterDiskANN,UNG}.

\vspace{-1mm}
\begin{definition}[Filtered Approximate Nearest Neighbour Search, FANNS]
\label{def:fanns}
Let \(\mathcal{D}\subset\mathbb{R}^d\) be a base dataset of \(n\) vectors, where each vector \(v_i \in \mathcal{D}\) is associated with a label set \(f_i\). Let \(v_q\in\mathbb{R}^d\) be a query vector with its own label set \(f_q\). Given an integer \(k\ge1\) and a filter constraint \(\mathcal{S}\), the filtered approximate \(k\)-nearest neighbour search algorithm aims to find  
the approximate $k$-NN set of size $k$,
\(
\mathcal{R}_{\mathcal{S}}(v_q)\;=\;\{x_1,\dots,x_k\}\;\subseteq\;\mathcal{D},
\)
such that two conditions are met:
\begin{enumerate}[leftmargin=*]
    \item \textbf{Filter Satisfaction}: For every vector \(x_j \in \mathcal{R}_{\mathcal{S}}(v_q)\), its corresponding label set \(f_j\) must satisfy the constraint \(\mathcal{S}\) with respect to the query's label set \(f_q\), denoted as \(f_j \mid_{\mathcal{S}} f_q\).
    \item \textbf{Approximate Proximity}: The set \(\mathcal{R}_{\mathcal{S}}(v_q)\) contains vectors that are an approximation of \(Gt_{\mathcal{S}}(v_q)\), that is, the true \(k\)-nearest neighbors to \(v_q\) among all vectors in \(\mathcal{D}\) that satisfy the filter constraint \(\mathcal{S}\).
\end{enumerate}
\end{definition}



Next, we present four filter constraints $\mathcal{S}$ commonly used in recent FANNS studies, followed by an illustrating example.

\vspace{-1mm}
\begin{definition}[Containment]
    A vector \(x_j \in \mathcal{R}_{\mathcal{S}}(v_q)\) with label set \(f_j\) satisfies \(f_j \mid_{\mathcal{S}} f_q\) iff $f_q \subseteq f_j$.
\end{definition}
\vspace{-2.5mm}
\begin{definition}[Overlap]
    A vector \(x_j \in \mathcal{R}_{\mathcal{S}}(v_q)\) with label set \(f_j\) satisfies \(f_j \mid_{\mathcal{S}} f_q\) if and only if $f_q \cap f_x \neq \emptyset$.
\end{definition}
\vspace{-2.5mm}
\begin{definition}[Equality]
    A vector \(x_j \in \mathcal{R}_{\mathcal{S}}(v_q)\) with label set \(f_j\) satisfies \(f_j \mid_{\mathcal{S}} f_q\) iff $f_q = f_j$.
\end{definition}
\vspace{-2.5mm}
\begin{definition}[Fixed-Length Equality]
    All vectors in dataset $\mathcal{D}$ must have the same numbers of labels, i.e., $|f_i|=|f_j|$ for $x_i, x_j \in \mathcal{D}$.
    And the filter constraint \(f_j \mid_{\mathcal{S}} f_q\) is satisfied if and only if $f_q = f_j$.
\end{definition}
\vspace{-2.5mm}

\begin{example}
    Figure~\ref{fig: filterdemo} illustrates how filtering scenarios affect nearest-neighbor results for identical queries. Without filtering, $v_5$ is the 1-NN (the nearest neighbour to query). Under containment constraints that desire vectors with label set containing query labels (i.e., $f_q \subseteq f_i$), $v_4$, $v_5$, and $v_6$ are excluded, as their label set does not contain $f_q$, making $v_3$ the 1-NN. Similarly, For equality constraints that require $f_i=f_q$, only $v_3$ and $v_7$ meet the filter condition, 
    leaving $v_3$ as 1-NN. Under overlap constraints that desire vectors with labels overlapped with query labels (i.e., $f_i \cap f_q \neq \emptyset$), only $v_4$ is excluded ($f_4\cap f_q = \emptyset$), restoring $v_5$ as 1-NN.
\end{example}
\vspace{-1mm}

Notably, beyond these filter constraints on categorical labels, 
recently \textit{range filters} also arise, where 
each vector $v_i$ has a single numerical attribute like a timestamp or a price. A query consists of a vector $v_q$ and a continuous range $[l, r]$, aiming to find the $k$-nearest neighbors to $v_q$ whose attributes fall within this range. Though there are a few specialized solutions~\cite{arkgraph, serf, unify}, we find that they are merely compatible with the majority of FANNS algorithms and thus are not covered in this survey.

\vspace{-1mm}
\subsection{Evaluation Metrics}
\label{sec:evaluation_metrics}

To assess the performance of FANNS algorithms, we evaluate both search accuracy and operational efficiency. These aspects are captured by the following standard metrics.

\vspace{-1mm}
\begin{definition}[Recall]
\label{def:recall}
Recall measures the quality of the search result by quantifying the fraction of true nearest neighbors that are successfully retrieved. For a given FANNS query \(v_q\) and a target number \(k\), the recall is defined as:
\vspace{-.5mm}
\[
\text{Recall@k} = \frac{|\mathcal{R}_{\mathcal{S}}(v_q) \cap Gt_{\mathcal{S}}(v_q)|}{|Gt_{\mathcal{S}}(v_q)|}, 
\]

\vspace{-1mm}
\noindent where \(Gt_{\mathcal{S}}(v_q)\) is the ground truth set of the FANNS query and \(\mathcal{R}_{\mathcal{S}}(v_q)\) is the result set returned by the algorithm.
\end{definition}
\vspace{-1mm}

To evaluate search efficiency, queries per second are commonly used to report query throughput.

\vspace{-1mm}
\begin{definition}[Queries Per Second, QPS]
\label{def:qps}
QPS is calculated as the total number of queries processed divided by the total time elapsed, i.e., how many queries the system can handle per unit time.
\end{definition}
\vspace{-1mm}

The performance of a FANNS algorithm, in terms of both recall and QPS, is heavily influenced by the difficulty of the query itself. The filter selectivity is a key factor influencing its difficulty.

\vspace{-1mm}
\begin{definition}[Filter Selectivity]
\label{def:selectivity}
Filter selectivity, denoted by \(\sigma_{\mathcal{S}}(f_q)\), is the fraction of the dataset that satisfies the given filter constraint \(\mathcal{S}\) with respect to a query's label set \(f_q\):
\vspace{-1mm}
\[
\sigma_{\mathcal{S}}(f_q) = \frac{|\{v_i \in \mathcal{D} \mid f_i \mid_{\mathcal{S}} f_q\}|}{|\mathcal{D}|}.
\]
\end{definition}

\section{
FANNS Algorithms}

\label{sec: overview of filter anns}

We categorize FANNS algorithms into three distinct classes according to the execution stage of the filtering process: filter-then-search, search-then-filter, and hybrid-search. Each category exhibits different search workflows. Subsequently, we will explain 
all specific algorithms for each class, and provide a detailed analysis
of their variations in index construction and query processing. 
Table \ref{tab:predicate_support} summarizes the support of various search algorithms for five key logical predicates, as well as their respective filter type classifications. 

\begin{figure}[t]
    \centering
    \includegraphics[width=\linewidth]{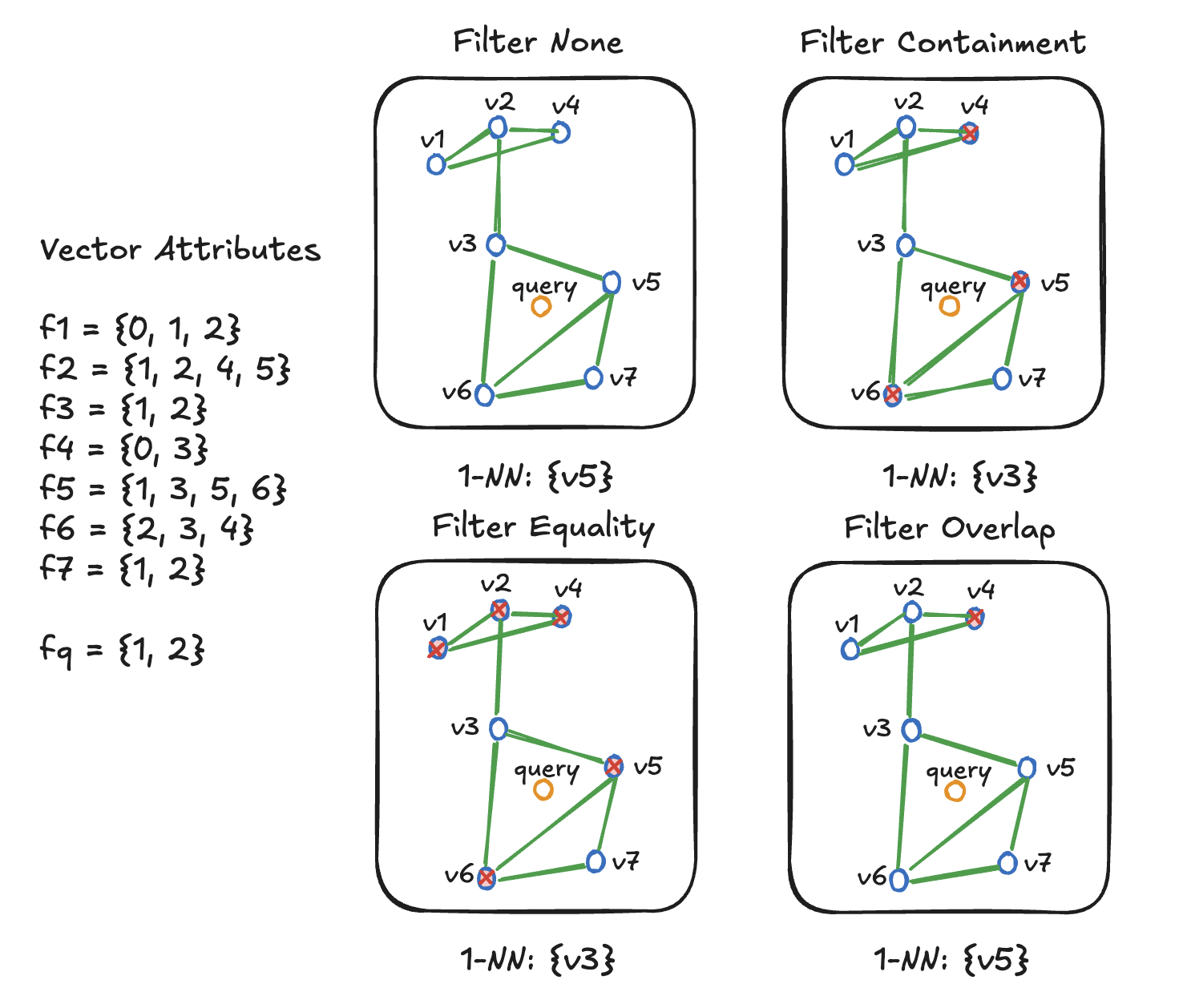}
    \vspace{-4mm}
    \caption{Visualization of FANNS: for 1-NN search (i.e., searching for the point closest to the query) under four scenarios (None, Containment, Equality, Overlap), results vary significantly across different scenarios.}
    \label{fig: filterdemo}
    \vspace{-2mm}
\end{figure}

\vspace{-1mm}
\subsection{Filter-Then-Search ANNS Algorithms}
\label{sec: Filter-then-search anns algorithm}

Filter-then-search algorithms explicitly or implicitly filter the entire dataset $\mathcal{D} = \{ \mathbf{x}_1, \mathbf{x}_2, \dots, \mathbf{x}_n \}$ to extract vectors satisfying specified label constraints. This process generates a filtered subset $\mathcal{D}_f = \{ \mathbf{x}_{k_1}, \mathbf{x}_{k_2}, \dots, \mathbf{x}_{k_m} \}$ where $\mathcal{D}_f \subseteq \mathcal{D}$. Subsequent nearest neighbor search operates exclusively within this filtered subset $\mathcal{D}_f$, requiring no additional constraint verification during query processing. The search procedure reduces to solving a conventional ANNS problem over the constrained dataset.

A critical implementation combines pre-filtering with vector quantization~\cite{elastic2025elasticsearch}. This approach first utilizes a fast bitset to identify the IDs of all points that satisfy the metadata constraints. Then, a quantization algorithm is used to rapidly compare the approximate distances between these candidate points and the query vector. Finally, a small subset of the most promising candidates is selected for a reranking step using their full-precision vector representations. 

Creating the filter map relies on a pre-computed inverted index, where each label maps to a bitset indicating all vectors that possess it. 
Given a query label set $f_q$, the filter map for \textit{Containment Scenario} is built by performing a bitwise AND operation on all the bitsets corresponding to the labels in $f_q$.
Conversely, the filter map for \textit{Overlap Scenario} is built by performing a bitwise OR operation on all these bitsets.
For \textit{Equality Scenario}, we can first perform a bitwise AND operation identical to the \textit{Containment Scenario}, followed by iterating these candidates to perform the more expensive check for exact set equality.
Given the machine word size $w$, such implementation achieves a time complexity of $O( |\overline{f_q}| \cdot N/w)$ for each query with an average of $|\overline{f_q}|$ bitset operations on a dataset of size $N$.
To store the bitset of each label, it requires $O(|f_{total}| \cdot N/w)$ space, where $|f_{total}|$ is the possible number of labels.



\vspace{-1mm}
\begin{example}
\normalfont
As shown in Figure~\ref{fig: filterdemo}, for a query $L_q = \{1, 2\}$ against a database of 7 items, we first retrieve the pre-computed bitsets. The set of items possessing label 1 is $\{v_1, v_2, v_3, v_5, v_7\}$, which corresponds to the bitset $\mathcal{B}_1 = \texttt{[1,1,1,0,1,0,1]}$. Similarly, the items with label 2 are $\{v_1, v_2, v_3, v_6, v_7\}$, yielding the bitset $\mathcal{B}_2 = \texttt{[1,1,1,0,0,1,1]}$. To find items that contain both labels (AND logic), we perform a bitwise AND: $\mathcal{B}_1 \land \mathcal{B}_2 = \texttt{[1,1,1,0,0,0,1]}$. To find items that contain either label (OR logic), we perform a bitwise OR: $\mathcal{B}_1 \lor \mathcal{B}_2 = \texttt{[1,1,1,0,1,1,1]}$.
\end{example}
\vspace{-1mm}

After pre-filtering with bitwise operations, the vector quantization is used for fast distance computing.
It has evolved from foundational techniques like Product Quantization (PQ)~\cite{PQ} and its Optimized Product Quantization (OPQ)~\cite{OPQ} variant to more recent algorithms like RaBitQ~\cite{rabitq} and Extended-RaBitQ~\cite{extended-rabitq}, which have been proven to have superior theoretical error bounds. 

\textbf{Pre-filter Brute Force.} We implement a baseline algorithm that first filters the dataset and then applies brute-force search. 
The filter map is used to retrieve the subset of database ($\mathcal{D}_f \subseteq \mathcal{D}$) that satisfies the filter constraints. Then, it performs a brute-force, full-precision search within filtered vector set ($\mathcal{D}_f$) and returns the top-$k$ results.

\begin{figure}[t]
    \centering
    \includegraphics[width=\linewidth]{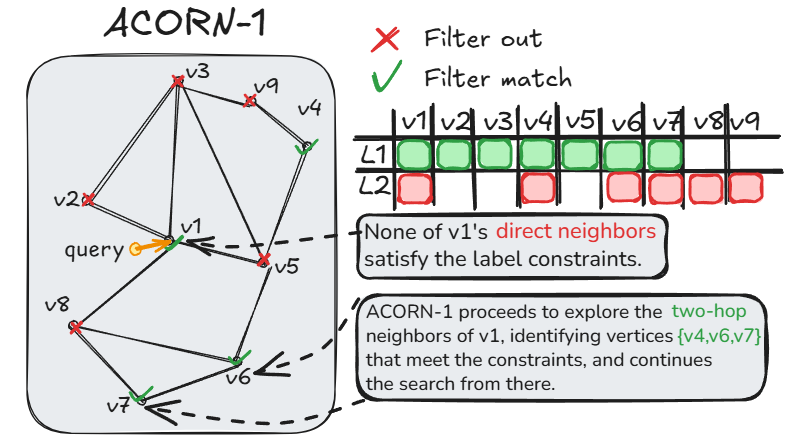}
    \vspace{-3mm}
    \caption{Example of ACORN-1: it illustrates that ACORN increases the breadth of search by finding two-hop neighbors.}
    \label{fig: acorndemo}
    \vspace{-3mm}
\end{figure}

\textbf{ACORN-$\gamma$~\cite{acorn}.} The ACORN-\(\gamma\) algorithm constructs its 
without any label information in advance. This allows for flexible data updates, as labels can be assigned or modified after the index is built. The core construction idea is to create a denser graph by expanding each node's degree by a factor \(\gamma\), which is determined based on an estimated selectivity. A specialized predicate-agnostic pruning technique is then applied to control index size. At query time, ACORN-\(\gamma\) first performs a full-dataset scan to generate a bitset of all vectors satisfying the filter constraint (which can be as complex as a regular expression). The subsequent graph traversal is then guided by this bitset. When expanding its candidate lists, it decompresses the neighbor information and checks filtering conditions, which broadens the search's scope. This algorithm can be combined with mainstream indexes like HNSW and DiskANN and performs well with high selectivity, and many vector search engines have added support for it, like ElasticSearch~\cite{elastic2025elasticsearch}. Due to this initial dataset-wide filtering step, we classify ACORN-\(\gamma\) as a filter-then-search approach.

\textbf{ACORN-1~\cite{acorn}.} In contrast to ACORN-\(\gamma\), ACORN-1(see Figure \ref{fig: acorndemo}) offers a lightweight alternative that aims to approximate the search performance of ACORN-\(\gamma\) while significantly reducing index construction time and memory cost. The key difference lies in its approach to neighbor expansion. Instead of creating a denser graph during construction, ACORN-1 builds a standard, un-pruned HNSW index. During the greedy search, then visiting a node \(v\), ACORN-1 dynamically expands the search scope to consider not only its direct (one-hop) neighbors but also its two-hop neighbors. This expanded candidate set is then filtered according to the query predicate before the best neighbors are pushed into 
the priority queue.

\begin{figure}[t]
    \centering
    \includegraphics[width=0.55\linewidth]{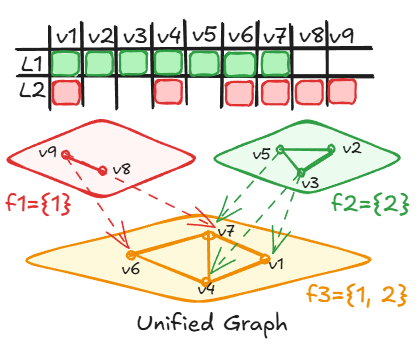}
    \vspace{-2mm}
    \caption{Example of UNG. The dataset is partitioned into disjoint groups by label sets: $f_1$=\{1\}, $f_2$=\{2\}, and $f_3$=\{1,2\}. Each group has its own intra-group graph with no overlap of vectors across groups. Inter-group connections follow the label-navigating graph and respect containment (e.g., \{1\}$\rightarrow$\{1,2\} and \{2\}$\rightarrow$\{1,2\}), so any edge $(v_i,v_j)$ implies $f_i\subseteq f_j$. }
    \label{fig: ungdemo}
    \vspace{-1mm}
\end{figure}

\textbf{UNG~\cite{UNG}.} The Unified Navigating Graph (UNG) algorithm pre-compiles label information into a label navigating graph structure during offline construction, which encodes the containment relationships of different label sets (see Figure \ref{fig: ungdemo}). It groups vectors by their labels, builds intra-group graphs, and adds inter-group edges based on a label navigation graph (LNG). This design guarantees that if an edge \((v_i, v_j)\) exists, their labels satisfy a containment relationship (\(f_i \subseteq f_j\)). 
By locating the entry labels that follow the filter constraints, greedy traversal on the proximity graph never requires further label checking,
making UNG a highly efficient filter-then-search method for containment and equality predicates. For equality scenarios, this index degenerates into a search on multiple independent small graphs, and for overlap scenarios, multiple rounds of searching are required, followed by merging the results to ensure recall in their implementation. However, its performance suffers significantly on queries with more complex constraints, such as label overlap 
scenarios, as multiple rounds of search are performed starting from different entries.

\subsection{Search-Then-Filter ANNS Algorithms}

\label{sec: Search-then-filter anns algorithm}

Search-then-filter algorithms perform an unconstrained approximate nearest neighbor search over the entire dataset $\mathcal{D} = \{ \mathbf{x}_1, \mathbf{x}_2, \dots,$ $ \mathbf{x}_n \}$, $\mathcal{R} = \{ \mathbf{x}_{r_1}, \mathbf{x}_{r_2}, \dots, \mathbf{x}_{r_l} \}$ ordered by their distance to the query,
where typically $l \gg k$ and $k$ denotes the number of search targets (i.e., Recall@$k$). This approach provides native compatibility with most existing ANNS algorithms. After performing the traditional ANNS search, the constraint verification proceeds as follows:
\begin{itemize}[leftmargin=*]
    \item If $\mathcal{R}$ contains at least $k$ vectors satisfying the label constraints, the top-$k$ valid results are returned immediately.
    \item Otherwise, the search scope $l$ is iteratively expanded to retrieve more candidates, repeating the verification process until $k$ valid results are obtained.
\end{itemize}

\begin{figure}[t]
    \centering
    \includegraphics[width=\linewidth]{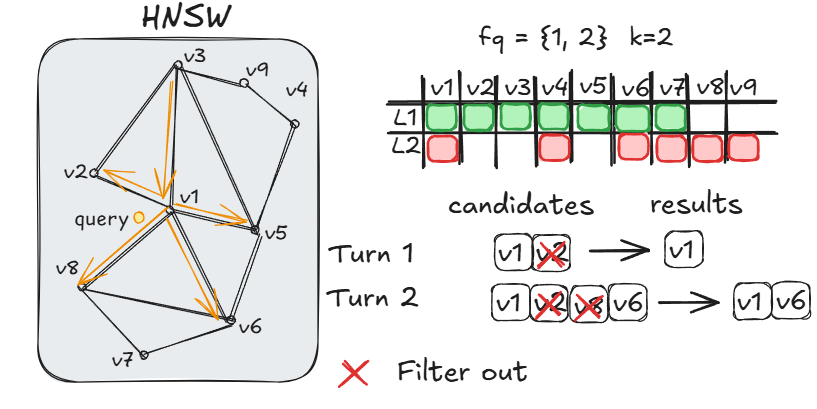}
    \vspace{-7mm}
    \caption{Example of Post Filter HNSW.}
    \label{fig: hnswdemo}
    \vspace{-3mm}
\end{figure}

\textbf{Post-filter HNSW~\cite{HNSW}.} We implement this baseline by combining the standard HNSW algorithm with a post-filtering strategy.(see Figure \ref{fig: hnswdemo}) First, a conventional HNSW index is built on the entire dataset \(\mathcal{D}\) without considering any label information. During a query, we perform an initial ANNS search on this index with a search parameter $l$ that controls the size of results. This result queue is then scanned to check for filter satisfaction. If at least \(k\) valid results are found, the top-\(k\) among them are returned. If the count is insufficient, the search is re-issued with an expanded scope (e.g., by doubling \(l\)) and the process is repeated until \(k\) valid neighbors are collected or a maximum search limit is reached.

\begin{figure}[t]
    \centering
    \includegraphics[width=\linewidth]{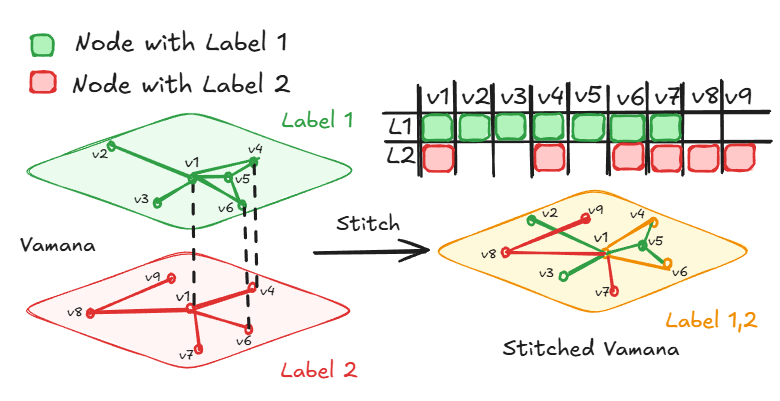}
    \vspace{-7mm}
    \caption{Example of Stitched DiskANN.}
    \label{fig: stitchdemo}
    \vspace{-2mm}
\end{figure}

\textbf{Post-filter IVF-PQ~\cite{ivfpq}.} Similarly, we implement another baseline using the Inverted File with Product Quantization (IVF-PQ) index. The post-filtering mechanism remains the same, but the underlying search scope is controlled by the $nprobe$ parameter, which specifies the number of inverted lists to visit.

\subsection{Hybrid-Search ANNS Algorithms}
\label{Hybrid-search anns algorithms}

Hybrid-search algorithms integrate label constraints directly into the search process, avoiding explicit pre-filtering or post-filtering stages. This approach typically employs some strategies, such as constraint neighbor expansion or distance fusion. They may build upon foundational ANNS methods, such as Filtered-DiskANN~\cite{filterDiskANN} that extends DiskANN~\cite{diskann} by deeply integrating label constraints into both index construction and query processing.

\begin{figure*}[t]
    \centering
    \includegraphics[width=1\linewidth]{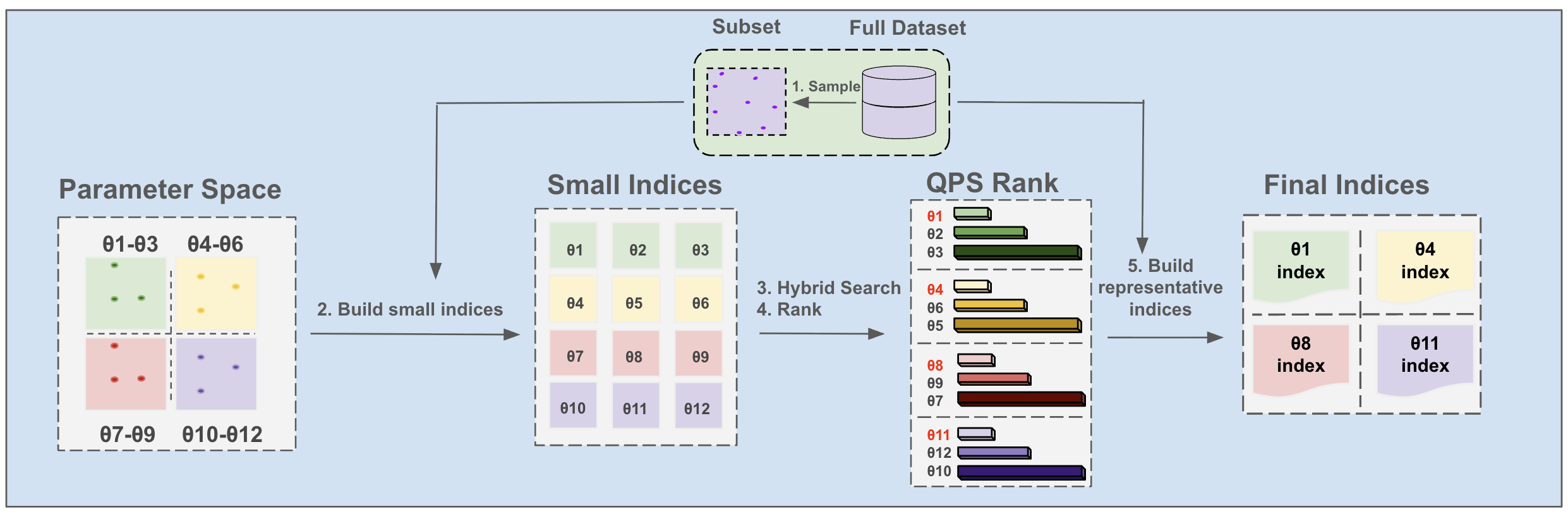}
    \vspace{-6mm}
    \caption{Illustration of Parameter Tuning Algorithm. The algorithm proceeds as follows: 1) randomly sample some points from the dataset as a subset (line~\ref{alg:line:sample}); 2) traverse the parameter space and build small indices (line~\ref{alg:line:build small indices}); 3) run hybrid searches for each small index (line~\ref{alg:line:hybrid search}); 4) perform parameter ranking in each uniformly divided subspace of the parameter space as detailed in line~\ref{alg:line:rank}.; 5) select one representative parameter from each subspace and construct full-scale indices (line~\ref{alg:line:select_best}).}
    \label{fig: tuning_demo}
    \vspace{-1mm}
\end{figure*}

\textbf{Filtered-DiskANN~\cite{filterDiskANN}.} This algorithm is based on the Vamana graph, the core graph index of DiskANN~\cite{Vamana}, but modifies both the construction and search processes to be label-aware. During index construction, when a new point is inserted, only neighbors that share at least one common label with the new point (i.e., \(f_i \cap f_q \neq \emptyset\)) are added to the candidate queue, and edges are then established with the candidates found along this search path. Finally, a label-aware pruning strategy is applied to refine the connections.  We note that the original implementation~\cite{diskann-github}  primarily supports singleton query label sets (\(|f_q| = 1\)). To accommodate more general cases (\(|f_q| > 1\)) and more scenarios, we extend the algorithm by adding an extra filtering step: after a node is popped from the candidate queue and before it is placed into the final result set, we perform a label check. Because this approach continually performs distance computations with label check throughout the search, we classify it as a hybrid-search algorithm.

\textbf{Stitched-DiskANN~\cite{filterDiskANN}.} In contrast to Filtered-DiskANN's graph approach, Stitched-DiskANN (see Figure \ref{fig: stitchdemo}) adopts a ``partition-then-stitch'' construction strategy. It first partitions the dataset by creating a separate Vamana subgraph~\cite{Vamana} for each individual label in the entire label set \(\mathcal{L}\). During the construction of each Vamana subgraph, a point is connected to other points whose labels intersect with its own; this design makes its search process highly suitable, as it only needs to expand neighbors that have label intersections with the query and ensures connectivity. These initially disconnected subgraphs are then ``stitched'' together into a single cohesive index. The stitching process leverages points that carry multiple labels, since these points naturally appear in multiple subgraphs, they serve as crucial bridge nodes that ensure connectivity.

\textbf{NHQ~\cite{NHQ}.} NHQ is a hybrid-search algorithm that builds upon foundational graph-based ANNS methods, such as kgraph~\cite{kgraph} and NSW~\cite{nsw1, nsw2}, and introduces modifications to better adapt to filtering scenarios. It integrates label constraints into both the graph construction and query process by fusing label-based distances with vector distance metrics. Specifically, NHQ defines a combined weight \(wt(v_i, v_j)\) using the Hamming distance of the labels between two nodes and the vector distance \(\delta(v_i, v_j)\), creating a weighted fusion of these two metrics. For graph construction, NHQ employs a Navigable PG (NPG) strategy, where edges are proactively established between points in diverse directions. This ensures better navigability during the search process, allowing queries to traverse the graph efficiently. In experimental evaluations, we use the NHQ-NPG-kgraph implementation, which incorporates these strategies for both index construction and query execution.

\textbf{CAPS \cite{CAPS}.} CAPS algorithm constructs a two-level index by first partitioning the dataset into coarse spatial clusters using k-means. 
Each cluster is then further subdivided into a hierarchy of fine-grained, label-based sub-clusters. 
This is done by iteratively creating sub-clusters for the most frequent label among the remaining vectors, up to a specified number of sub-clusters \(h\). 
At query time, CAPS first identifies the nearest coarse cluster centroid. 
It then sequentially probes the second-level sub-clusters within that cluster in their order of creation to find matching results.

\section{Parameter Tuning}
\label{sec: param tuning}

Existing studies demonstrate that vector databases exhibit a high-dimensional parameter space where configurations are highly interdependent \cite{vdtuner}. This strong parameter coupling prevents independent optimization of individual parameters. Furthermore, tuning constitutes a complex two-objective optimization problem (maximizing both QPS and recall), making it challenging to identify parameter configurations that yield optimal trade-offs in the QPS-Recall space.

To address these challenges, 
we propose a parameter tuning approach as shown in Figure~\ref{fig: tuning_demo}, which is designed for:

\begin{itemize}[leftmargin=*]
\item \textbf{Coverage:} Comprehensive exploration of the parameter space.
\item \textbf{Balance:} Optimal trade-offs via the Pareto frontier.
\item \textbf{Robustness:} Near-optimal parameters for each specific context.
\end{itemize}

Algorithm \ref{alg:parameter_tuning} begins by partitioning the entire parameter space $\Omega$ into smaller, manageable subspaces (line~\ref{alg:line:partition}). 
This strategy ensures a broad exploration of different parameter regions. 
To accelerate the evaluation process, we perform searches on a randomly sampled subset of the dataset $\mathcal{D}'$ (line~\ref{alg:line:sample}).
For each parameter configuration $\theta$ within a subspace, we evaluate its performance across all defined search scenarios and compute the average QPS and Recall (line~\ref{alg:line:avg_metrics}).

\begin{table*}[t]
\centering
\caption{Statistics of Datasets}
\label{tab:datasets}
\vspace{-3mm}
\setlength{\tabcolsep}{4.6mm}{
\begin{tabular}{l l c c l l}
\toprule
\textbf{Dataset} & \textbf{Type} & \textbf{Dim} & \textbf{Number of Vectors} & \textbf{Original Labels} & \textbf{Application} \\
\midrule
arXiv & text & 768 & 132,678 & year, month, task, etc & Academic retrieval \\
TripClick & text & 768 & 1,055,976 & clinical area & Health web search \\
LAION1M & image & 512 & 1,000,448 & entities, locations, etc & Image retrieval \\
YFCC & image+audio & 192 & 1,000,000 & classes & Image retrieval \\
YouTube Audio & audio & 128 & 5,000,000 & classes & Audio retrieval \\
YouTube Video & audio & 1024 & 1,000,000 & classes & Video retrieval \\
\bottomrule
\end{tabular}
}
\end{table*}

A key step in our method is to normalize the comparison of different parameter sets. 
We evaluate how efficiently a configuration can achieve specific recall targets. 
We define a set of fixed recall levels (e.g., 0.8, 0.9, 0.95) (line~\ref{alg:line:recall_targets}) and use linear interpolation to estimate the QPS that each configuration $\theta$ would achieve at these exact recall values (line~\ref{alg:line:interpolation}). 
Each configuration is then assigned a rank based on its interpolated QPS values (line~\ref{alg:line:rank}). Finally, for each subspace, we select the parameter configuration with the best overall rank as the representative for that region.

\begin{algorithm}[t]
\small
\caption{Parameter Tuning Algorithm}
\label{alg:parameter_tuning}
\begin{algorithmic}[1] 
\Require Algorithm $\mathcal{A}$, Parameter space $\Omega$, Dataset $\mathcal{D}$, Queryset $\mathcal{Q}$
\Ensure A set of representative parameters

\State Divide the parameter space $\Omega$ into subspaces $\{\Omega_1, \Omega_2, \dots, \Omega_n\}$. \label{alg:line:partition}
\State $Result \gets \emptyset$
\For{each subspace $\Omega_i \in \{\Omega_1, \Omega_2, \dots, \Omega_n\}$}
    \State Initialize $\Theta_i \gets \emptyset$
    \State Generate a sampled dataset $\mathcal{D}' \gets$ \Call{RandomSample}{$\mathcal{D}$} \label{alg:line:sample}
    \For{each parameter set $\theta \in \Omega_i$}
        \State Initialize $\{qps_s, recall_s\} \gets \emptyset$ for all scenarios $s \in \mathcal{S}$
        \State $\mathcal{I}\gets $ \Call{ConstructIndex}{$\mathcal{A}, \theta, \mathcal{D}$} \label{alg:line:build small indices}
        \For{each scenario $s \in \mathcal{S}$}
            \State $\{qps_s, recall_s\} \gets$ \Call{PerformSearch}{$\mathcal{A}, \mathcal{I}, s, \mathcal{Q}$} \label{alg:line:hybrid search}
        \EndFor
        \State Average metrics: $\{\overline{qps}, \overline{recall}\} \gets \text{avg}_s(\{qps_s, recall_s\})$ \label{alg:line:avg_metrics}
        \State Define recall targets $\mathcal{R} \gets [\text{recall}_1, \text{recall}_2, \text{recall}_3]$ \label{alg:line:recall_targets}
        \State $\text{qps}_1, \text{qps}_2, \text{qps}_3 \gets$ \Call{Interpolation}{$\{\overline{qps}, \overline{recall}\}, \mathcal{R}$} \label{alg:line:interpolation}
        \State Rank $\theta$ based on $\text{qps}_1, \text{qps}_2, \text{qps}_3$: $r_\theta$ \label{alg:line:rank}
        \State $\Theta_i \gets \Theta_i \cup \{(\theta, r_\theta)\}$
    \EndFor
    \State $\theta^* \gets \arg\min_{\theta \in \Theta_i} r_\theta$ \label{alg:line:select_best}
    \State $Result \gets Result \cup \{\theta^*\}$
\EndFor
\State \Return $Result$
\end{algorithmic}
\end{algorithm}

\section{Experiments}

In this section, we systematically evaluate diverse real-world workloads within a unified benchmark, thoroughly comparing FANNS algorithms under parameter-aware settings.

\subsection{Experimental Setup}
\label{sec: experiment setup}

\begin{figure*}[t]
    \centering
     \vspace{-3mm}
    \includegraphics[width=1\linewidth]{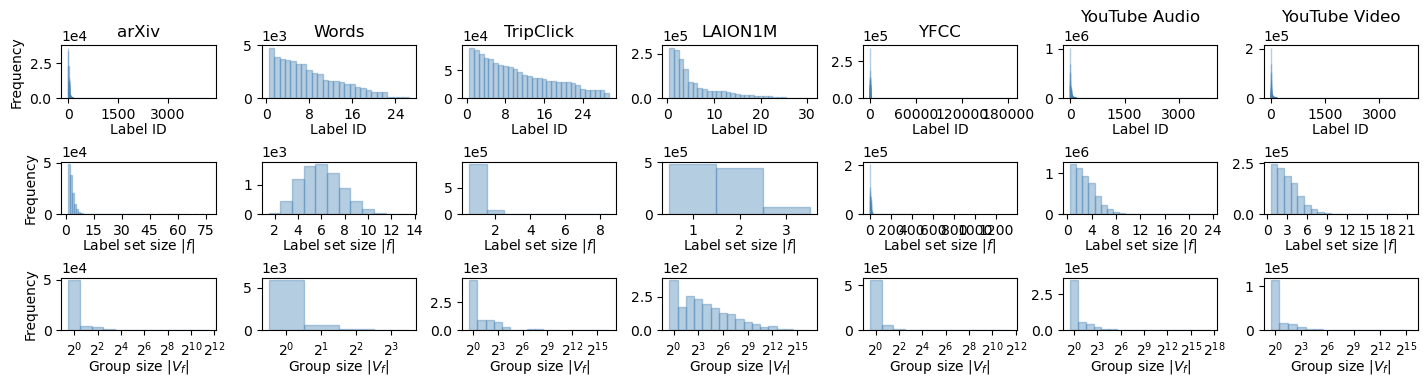}
    \vspace{-6mm}
    \caption{Distributions of base label, size of base label sets, and the number of groups.}
    \label{fig: label_stats}
    \vspace{-1mm}
\end{figure*}

\noindent \textbf{Scenarios.}
As presented in Section~\ref{sec: problem definition}, we cover four filtering scenarios to comprehensively evaluate existing FANNS solutions, including \textit{Containment}, \textit{Overlap}, \textit{Equality}, and \textit{Fixed-Length Equality}.
%
%
The majority of methods, such as ACORN~\cite{acorn} and Post-Filter HNSW~\cite{HNSW}, natively support the first three scenarios (Containment, Equality, and Overlap), while others, like FilteredDiskANN~\cite{filterDiskANN}, can be adapted with minor modifications. In contrast, a few algorithms, such as NHQ and CAPS~\cite{CAPS}, are specifically designed for the more restrictive \textit{Fixed-Length Equality Scenario}, which is 
a specialized subset of the broader \textit{Equality Scenario}.

\noindent\textbf{Datasets.}
Table~\ref{tab:datasets} summarises $6$ real-world datasets used in the experiments. All selected datasets include real-world labels. Some are provided directly in vector format, while others contain raw images or text. For the latter, we generated embeddings using pretrained models. Each dataset will be described in detail below:

\vspace{-1mm}
\begin{itemize}[leftmargin=*] 

\item \textbf{arXiv} \cite{Arxiv}. 
We use the same dataset as \cite{UNG}: 132,687 papers with methodology embeddings from \cite{arxivembedding}. 
    Label-equality uses $26$ distinct labels (12 months + 14 years). 
    Other scenarios use attributes: task (1678 values), method (890), dataset (1637), with possible multi-values per vector.


\item \textbf{TripClick} \cite{tripclick} A large-scale health information retrieval dataset, comprising click logs from the Trip Database health web search engine. The dataset includes approximately $1$ million user interactions with medical field, clinical area list and publication date, etc. We selected the 27 most frequent labels and categorized all the remaining ones as others.

\item \textbf{LAION1M} \cite{LAION} A large-scale, publicly available dataset of 400 million CLIP-filtered\cite{CLIP} image-text pairs, where the natural language text captions serve as descriptive labels for the images. We only selected the first $1000448$ items from $4$ million embeddings.

\item \textbf{YFCC} \cite{YFCC}. A standard CV dataset with 100M image/video embeddings in total. Labels include metadata (identifiers, tags, dates, etc). We use precomputed embeddings and formatted labels from \cite{bigann}. We randomly selected $1$M points from the dataset.

\item \textbf{YouTube-Audio} and \textbf{YouTube-Video}~\cite{youtube} A large-scale benchmark of millions of videos annotated with thousands of machine-generated topical entities, providing distinct, pre-computed visual and audio features extracted at one-second intervals. The two datasets share the same attributes.

\end{itemize}

Figure~\ref{fig: label_stats} presents the label distribution statistics across all evaluated real-world datasets. The top row shows that the label ID frequencies typically follow a long-tailed distribution. The middle row indicates the number of labels associated with the data points. The bottom row, which illustrates the group size (the number of items sharing an identical set of labels), also demonstrates a skewed distribution. We ensure that our query labels are sampled to reflect these inherent distributions of the base labels by random sampling, thereby simulating a realistic retrieval environment.

\noindent \textbf{Metrics.}
We employ the \textit{QPS} (Queries Per Second) and \textit{Recall@k} metrics defined in Section~\ref{sec:evaluation_metrics}. Unless otherwise specified, $k=10$ is used as the default top-$k$ value. For all experiments, we ensure label selectivity defined in Section~\ref{sec:evaluation_metrics} thresholds guarantee at least $k$ ground-truth results per query.

\noindent \textbf{Experiment Environment.}
Programs are implemented in C++ and compiled with -Ofast Optimization. 
All the experiments are executed on a Linux server with Intel(R) E5-2596v4 CPU @2.2GHz and 220GB of RAM.
For all algorithms, we use $16$ threads to construct the index.
To process ANNS queries, we use $16$ threads for searching. Furthermore, for single-threaded algorithms~\cite{NHQ, CAPS}, we have added a multi-threaded implementation. The multithreading for all algorithms is achieved using the C++ OpenMP library~\cite{openmp08}. Due to the random access nature of node traversal~\cite{HNSW, NSG, graph_reorder, diskann++}, implementing parallelization within graph-based algorithms is not straightforward. Therefore, we only apply simple parallelization across different queries, similar to \cite{filterDiskANN, UNG}. We have implemented bitset calculation for Acorn as detailed in Section~\ref{sec: Filter-then-search anns algorithm}, which is significantly faster than its brute-force bitset calculation~\cite{ACORN-codes} and previous implementations~\cite{UNG}. This bitset is also applied to the post-filtering process in Faiss-HNSW and Faiss-IVFPQ~\cite{faiss}. Additionally, we have implemented multi-label search for FilterDiskANN and StitchedDiskANN~\cite{filterDiskANN}. We did not alter the graph construction process; instead, we check if the constraints for all labels are met when populating the result list.

\begin{table}[t]
\centering
\caption{Base Label Requirements for Index Construction}
\vspace{-2mm}
\label{tab:label_requirements}
\setlength{\tabcolsep}{8mm}{
\begin{tabular}{@{}cl@{}}
\toprule
\textbf{Base Label Requirement} & \textbf{Algorithms} \\ 
\midrule
\multirow{6}{*}{Requires Base Labels} & NHQ$^\dagger$~\cite{NHQ} \\
& CAPS~\cite{CAPS} \\
& UNG~\cite{UNG} \\
& Filtered-DiskANN$^\dagger$~\cite{filterDiskANN} \\
& Stitched-DiskANN$^\dagger$~\cite{filterDiskANN} \\
& Brute-Force~\cite{UNG} \\
\midrule
\multirow{4}{*}{Require No Base Labels} & ACORN-1~\cite{acorn} \\
& ACORN-$\gamma$~\cite{acorn} \\
& Post-Filter HNSW~\cite{HNSW} \\
& Post-Filter IVFPQ~\cite{ivfpq} \\
\bottomrule
\end{tabular}
}
\end{table}

\subsection{Performance of Constructing Index}

Before evaluating online search performance, we first measure offline index construction time and analyze parameters affecting graph-building efficiency. Since our largest dataset is limited to 500M vectors\cite{youtube}, the index construction cost critically determines algorithmic scalability to larger datasets. We also evaluate whether index building requires \textit{base label} information, which impacts update-friendliness when base labels change.

In Table~\ref{tab:label_requirements},
algorithms marked with $\dagger$ require base labels during initial index construction but support efficient incremental updates. For these methods, small-scale modifications (vector/label insertions, deletions, or updates) can be handled with minimal impact on overall index structure or search performance.

\begin{figure}[t]
    \centering
    \vspace{1mm}
    \includegraphics[width=\linewidth]{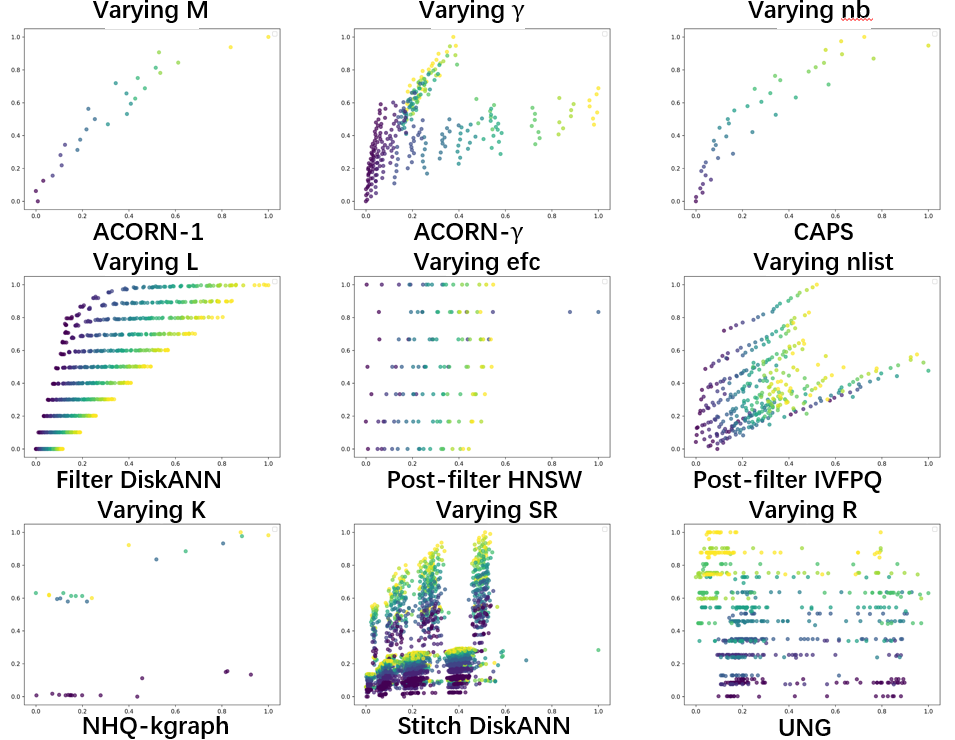}
    \vspace{-5mm}
    \caption{Scatter plots showing the effect of varying parameters on index time and size for different algorithms. Each point represents a specific parameter configuration, normalized to the [0, 1] range.}
    \label{fig: index_scatter}
    \vspace{-4mm}
\end{figure}

Figure~\ref{fig: index_scatter} illustrates the impact of varying parameters on the indexing time and index size for different algorithms. Each scatter plot corresponds to a specific algorithm, with the horizontal axis representing the normalized indexing time and the vertical axis representing the normalized index size. Different parameter values are encoded by the colors of the points. Notably, while each algorithm may have multiple tunable parameters, only one representative parameter was selected for visualization in each scatter plot.

\subsection{Overall Query Performance}
Our evaluation is conducted on 6 real-world datasets spanning diverse application domains, as detailed in~Section~\ref{sec: experiment setup}. We leverage the original labels provided with each dataset, which are preprocessed by mapping them to a dense, consecutive integer range. To ensure a fair and consistent evaluation across our three experimental scenarios, the query sets were curated so that each query has at least $10$ ground-truth items that fulfill the specified label constraint.

\begin{figure*}[t]
    \centering
    \includegraphics[width=1.0\linewidth]{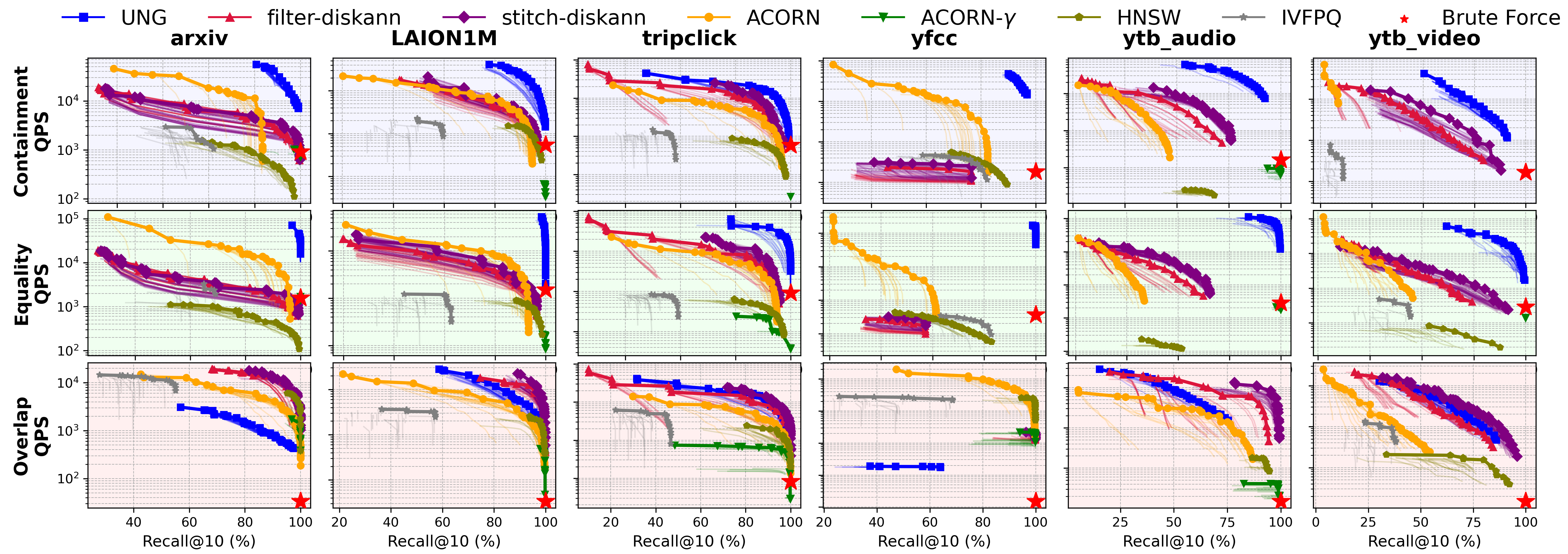}
    \vspace{-6mm}
    \caption{Query performance on $6$ real-world datasets.}
    \label{fig: basic experiment}
    \vspace{-1mm}
\end{figure*}

Figure~\ref{fig: basic experiment} shows the performance of all algorithms across various datasets, which have significant differences in dimensionality and the number of labels. The QPS-recall curve for each parameter set is represented by a light-colored line. For the points on the Pareto Frontier, they are bolded and shown in a darker color. Unless otherwise specified, this convention applies to all subsequent figures. 
Accordingly, the algorithms exhibit varied performance on these datasets. Overall, \texttt{UNG} is well-suited for the containment and equality scenarios due to its unique filter-then-search design. \texttt{DiskANN} performs well in the overlap scenario while maintaining a good balance in others. The specific factors influencing these performance differences will be discussed in detail in the following experimental sections; this section serves as a brief overview.


\subsection{Effect of Fixed Length Equality}
\label{sec: fixed length equality experiment}

In previous studies~\cite{NHQ, CAPS}, the Fixed Length Equality is a commonly used scenario. For this experiment, we selected two representative datasets, arXiv~\cite{Arxiv} and YFCC~\cite{YFCC}. Since this scenario is the simplest among the four, all algorithms can handle it without modifying their original implementations. Therefore, this experiment involves all the algorithms. As such structured labels are not part of the original datasets, we followed the implementation in NHQ paper~\cite{NHQ} to generate synthetic data with a fixed length of $4$. Each position in the label has three possible values, chosen with equal probability. Consequently, there are $3^4$ possible combinations, and the selectivity of each combination is $1/81$.

\begin{figure}[t]
    \centering
    \includegraphics[width=1\linewidth]{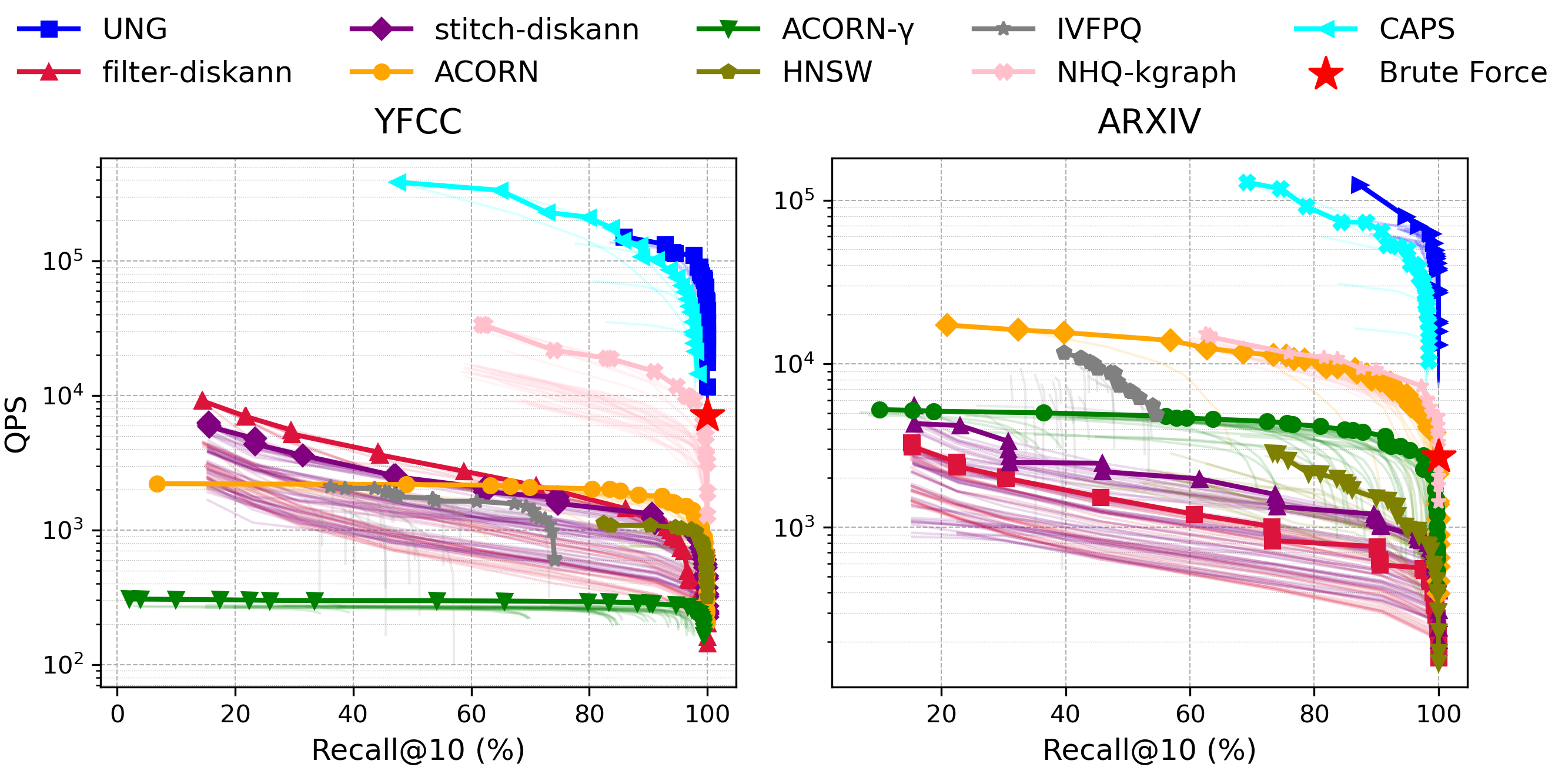}
    \vspace{-6mm}
    \caption{Result of Fixed Length Equality Scenario.}
    \label{fig: equal length experiment}
\end{figure}

Figure~\ref{fig: equal length experiment} shows the performance of $10$ algorithms in this experiment. Each algorithm has several optimal parameter combinations. In this scenario, most algorithms are capable of achieving high recall values by adjusting their search parameters. Among all tested methods, \texttt{UNG} and \texttt{CAPS} consistently deliver the highest QPS, surpassing $10^5$ on the YFCC dataset at high recall levels. The performance on the YFCC dataset shows a clear stratification of the algorithms, with distinct performance tiers. On the smaller, lower-dimensional arXiV dataset, most algorithms exhibit a noticeable improvement in QPS. This performance lift is particularly significant for \texttt{ACORN} and \texttt{ACORN-$\gamma$}.


\subsection{Effect of Varying Query Label Length}

\label{sec: query label length experiment}

\begin{figure*}[t]
    \centering
    \vspace{-3mm}
    \includegraphics[width=1.0\linewidth]{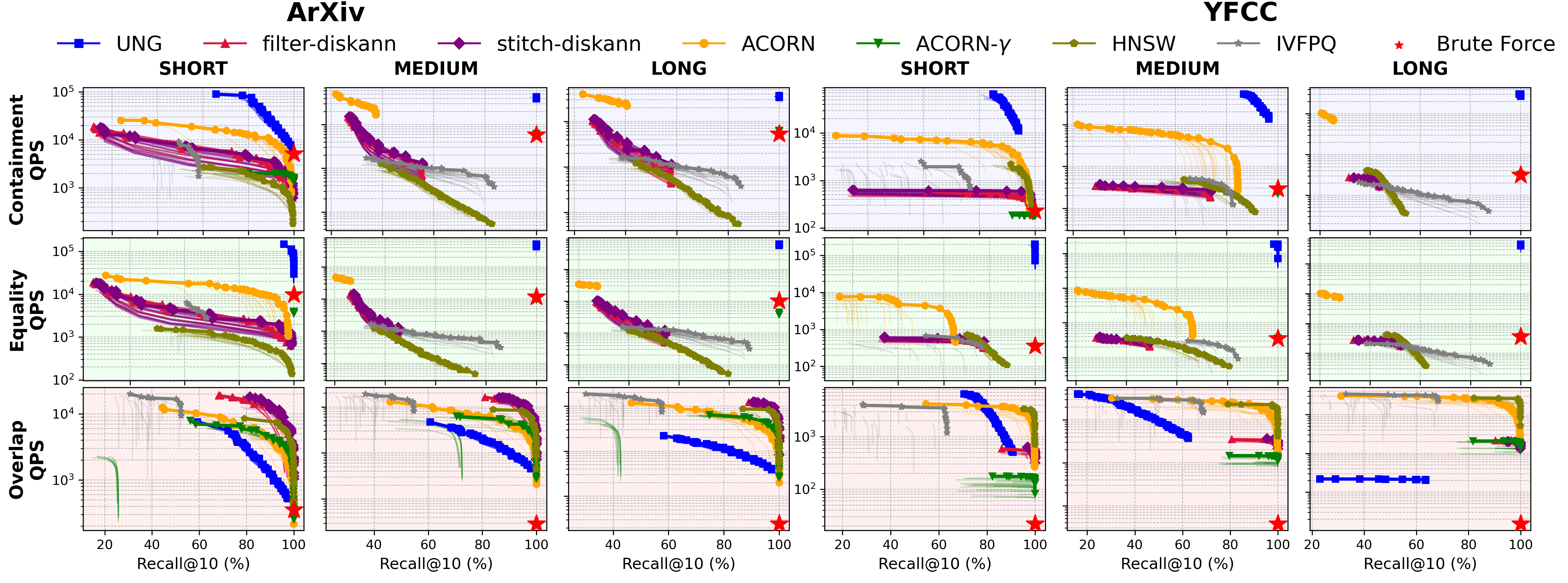}
    \vspace{-7mm}
    \caption{Results of Varying Query Label Length.}
    \label{fig: label length experiment}
    \vspace{-2mm}
\end{figure*}

\begin{figure*}[t]
    \centering
    \includegraphics[width=1.0\linewidth]{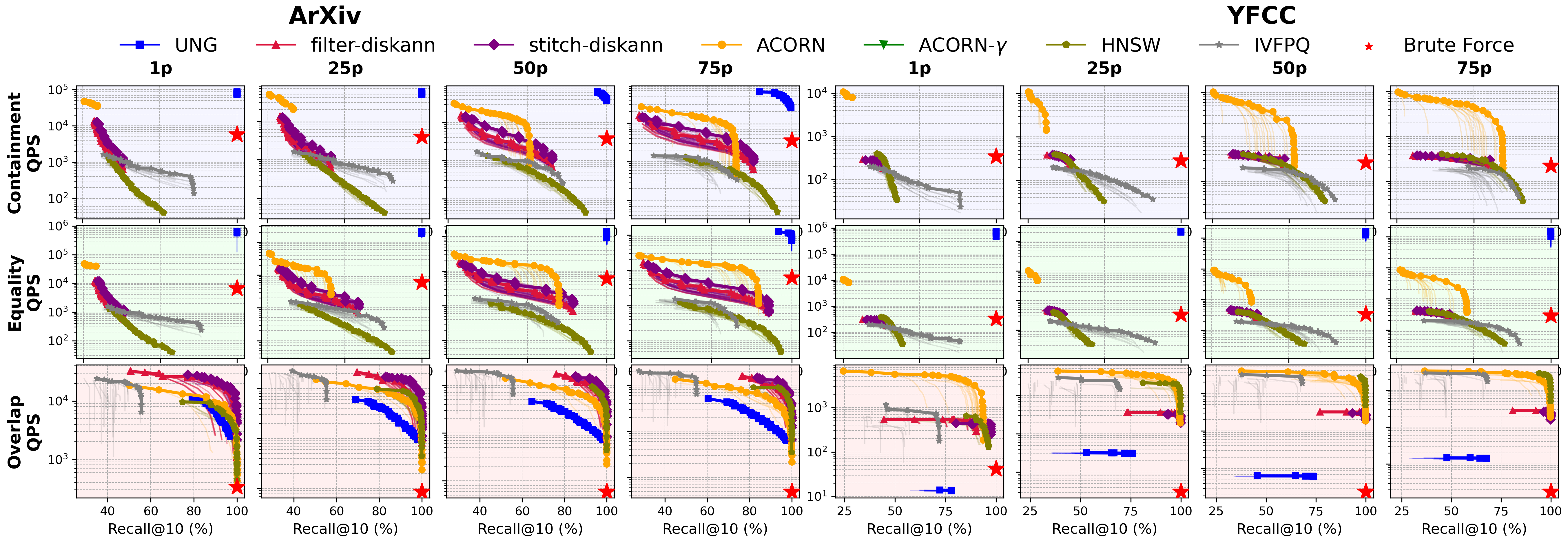}
    \vspace{-7mm}
    \caption{Results of Varying Query Selectivity.}
    \label{fig: selectivity experiment}
    \vspace{-1mm}
\end{figure*}

In all subsequent experiments, we use real-world label data. Similar to previous studies~\cite{UNG}, we conduct experiments on two representative datasets for three scenarios. For each scenario, we categorize the size of the query label $|f_q|$ into three groups: short length, medium length, and long length. From each group, we randomly select $1,000$ queries to explore the impact of query label length on the difficulty of the queries. Unless otherwise specified, all subsequent experiments are conducted on the two representative datasets, \textit{arxiv}~\cite{Arxiv} and \textit{yfcc}~\cite{YFCC}, and the three scenarios: \textit{containment}, \textit{equality}, and \textit{overlap}. Since the original implementations and code of NHQ~\cite{NHQcode} and CAPS~\cite{capscode} do not support these scenarios, the subsequent experiments will not include comparisons with these two methods.

When the query labels are short, nearly all algorithms can achieve a high recall range (97\%+) on both datasets. As shown in Figure~\ref{fig: label length experiment}, as the query label length increases, significant performance divergence in terms of recall emerges among the algorithms, accompanied by a general decrease in QPS.
In the \texttt{Containment} and \texttt{Equality} scenarios, UNG exhibits superior performance due to its specialized design. Counter-intuitively, its QPS tends to increase as the labels get longer in these specific cases. A pre-filter brute force approach also performs remarkably well in these scenarios; although it employs a full-precision search, the number of candidate points after filtering is small, thus limiting the number of distance computations and maintaining a relatively high QPS. Similarly, the performance of \texttt{ACORN-$\gamma$} often resembles the brute-force method, which can be attributed to cases where low selectivity effectively prunes the search space to a small number of candidates. In this scenario, as query labels grow longer, \texttt{ACORN-1} and \texttt{DiskANN} algorithms quickly hit a bottleneck and struggle to improve recall. This is because during their search queue expansion, if the points in their neighbors do not satisfy the label constraints, it becomes difficult to expand to more distant points that do meet those constraints.

In contrast, the performance trends are reversed in the \texttt{Overlap} scenario. Here, methods such as \texttt{filter-diskann}, \texttt{stitch-diskann}, \texttt{HNSW}, and the \texttt{ACORN} variants all have the potential to be optimal under certain conditions. Notably, in this scenario, both the QPS and recall of \texttt{UNG} degrade rapidly as the label length increases.

\subsection{Effect of Varying Query Selectivity}

\label{sec: query selectivity experiment}

Similar to the setup in previous works~\cite{filterDiskANN, UNG}, we group the real-world query labels based on their selectivity $\sigma_{\mathcal{S}}(f_q)$. According to the specific scenario, the queries are divided into four groups: 75th, 50th, 25th, and 1st percentiles. For each group, we evaluate the performance of different algorithms under each scenario.

In Figure \ref{fig: selectivity experiment}, as selectivity increases from left to right, most methods exhibit an increase in recall while maintaining their QPS. \texttt{UNG} and \texttt{ACORN} typically retain higher QPS over a broad recall range, whereas post-filtering algorithms tend to lose throughput more rapidly at low selectivity. In containment and oequality scenario, \texttt{UNG} requires searching only a very small subgraph, so it does not encounter nodes that violate label constraints; consequently, lower selectivity yields higher QPS and recall approaches 100\%.

In the overlap scenario, both \texttt{DiskANN} and \texttt{ACORN} achieve strong results because their neighbor selection retains only vectors with non-empty intersection with the query, which preserves recall. \texttt{UNG} shows no notable degradation on smaller datasets, but as the dataset grows it must re-rank many more candidates, causing both QPS and recall to drop.

\subsection{Effect of Varying Top-k}

\label{sec: topk experiment}

\begin{figure*}[t]
    \centering
    \includegraphics[width=1.0\linewidth]{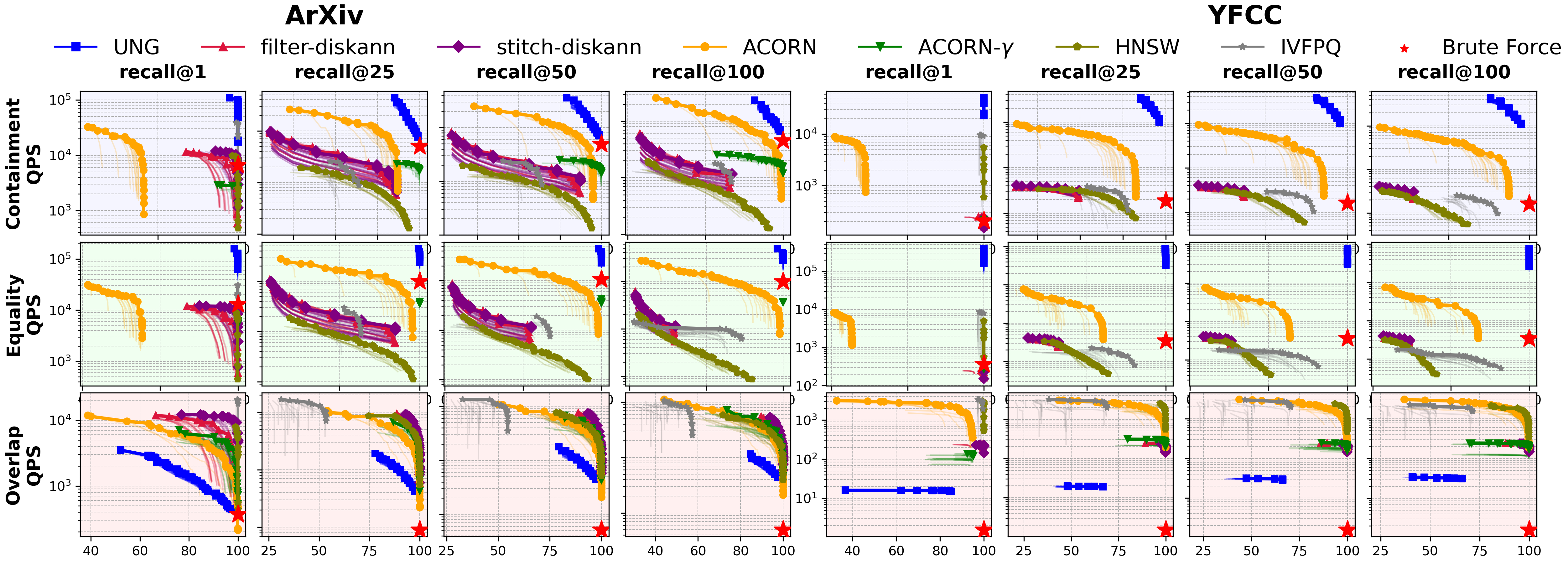}
    \vspace{-7mm}
    \caption{Results of Varying Top-k.}
    \label{fig: recall experiment}
    \vspace{-2mm}
\end{figure*}

\begin{figure*}[t]
    \centering
    \includegraphics[width=1.0\linewidth]{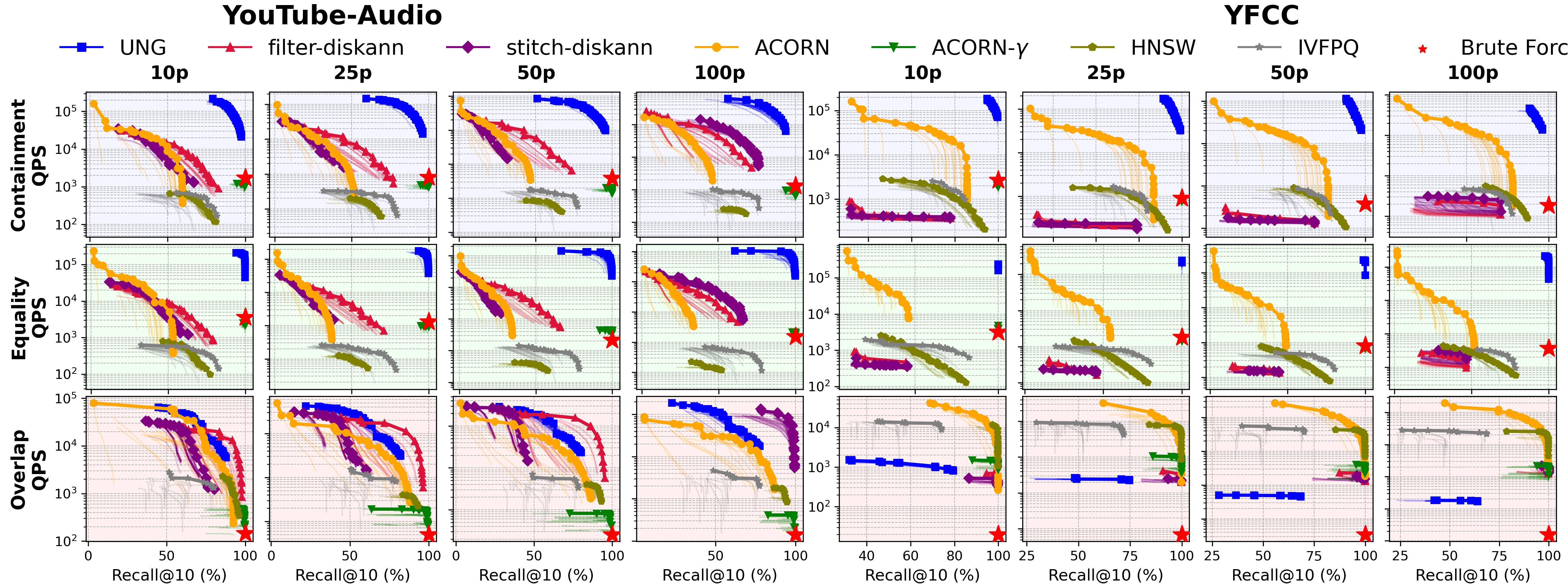}
    \vspace{-6mm}
    \caption{Results of Varying Base Dataset Size.}
    \label{fig: base experiment}
    \vspace{-1mm}
\end{figure*}

Many real-world scenarios, such as face recognition and retrieval-augmented generation (RAG), have special requirements for both the selection of top-k and its accuracy~\cite{RAGsurvey2023}. 
Thus, in this set of experiments, we evaluate the impact of varying the value of \textit{top-k} on the performance of different algorithms. For each scenario, we test a range of top-k values, including $k=1$, $k=25$, $k=50$, and $k=100$. 
We guarantee that the selected queries always have at least $k$ ground truth results, regardless of the value of $k$.

As shown in Figure \ref{fig: recall experiment}, across both the arXiv and YFCC datasets, the \texttt{UNG} and \texttt{ACORN} algorithm consistently demonstrates superior performance in the vast majority of scenarios. For instance, in the arXiv Containment scenario at recall@100, UNG reaches a QPS of $10^4$ while \texttt{ACORN} sustains a QPS of approximately $5*10^3$ while achieving over 95\% recall, whereas competing methods like stitch-diskann achieve lower QPS under the same conditions. noted that while \texttt{post-filtering IVFPQ} achieves high QPS and recall at \texttt{recall@1} across all scenarios, its performance deteriorates significantly as the value of $k$ in top-k increases. This is because the ground truth vectors can be scattered across different inverted buckets, making it very difficult to retrieve the complete set.

Similarly, the \texttt{UNG} algorithm demonstrates particularly poor performance in the Overlap scenario. In this context, while a considerable number of algorithms achieve comparable results on the arXiv dataset, their performance diverges significantly on the YFCC dataset, which features a larger number of labels. On the more challenging YFCC dataset, both \texttt{ACORN} and \texttt{Post-filter HNSW} stand out by maintaining exceptionally high and stable QPS and recall rates. Notably, their performance remains robust and does not degrade significantly even as the value of $k$ increases.

\subsection{Effect of Varying Base Dataset Size}

\label{sec: base experiment}

To simulate the performance of different algorithms under varying base dataset sizes, we used two relatively large datasets: \textit{yfcc}~\cite{YFCC} and \textit{youtube-audio}~\cite{youtube}. For each dataset, we selected subsets containing 10\%, 25\%, 50\%, and 100\% of the original data points. These subsets were then tested across three scenarios. We aim to analyze how the scalability of the algorithms is affected as the dataset grows.

In contrast to our earlier findings, increasing the dataset scale from $10\%$ to $100\%$ has a limited impact on performance. In Figure~\ref{fig: base experiment}, the majority of algorithms exhibit only a marginal decrease in QPS, while their recall ranges remain largely stable. Consequently, the QPS-recall curves for a given scenario show minimal variation across different scales, indicating that dataset size is not the primary performance bottleneck. Inherently difficult queries that fail on the smaller dataset continue to do so on the larger one.

The choice of dataset reveals more pronounced differences. On YouTube-Audio, \texttt{UNG} and \texttt{DiskANN} are the top performers. Conversely, on the YFCC dataset, which features more intricate label relationships, the performance landscape is more varied. In these scenarios, \texttt{UNG}, \texttt{ACORN}, and \texttt{HNSW} each demonstrate distinct advantages. We must note that this entire evaluation is based on in-memory indexes; disk-based counterparts were not investigated.

\begin{figure}[t]
    \centering
    \vspace{-1mm}
    \includegraphics[width=1\linewidth]{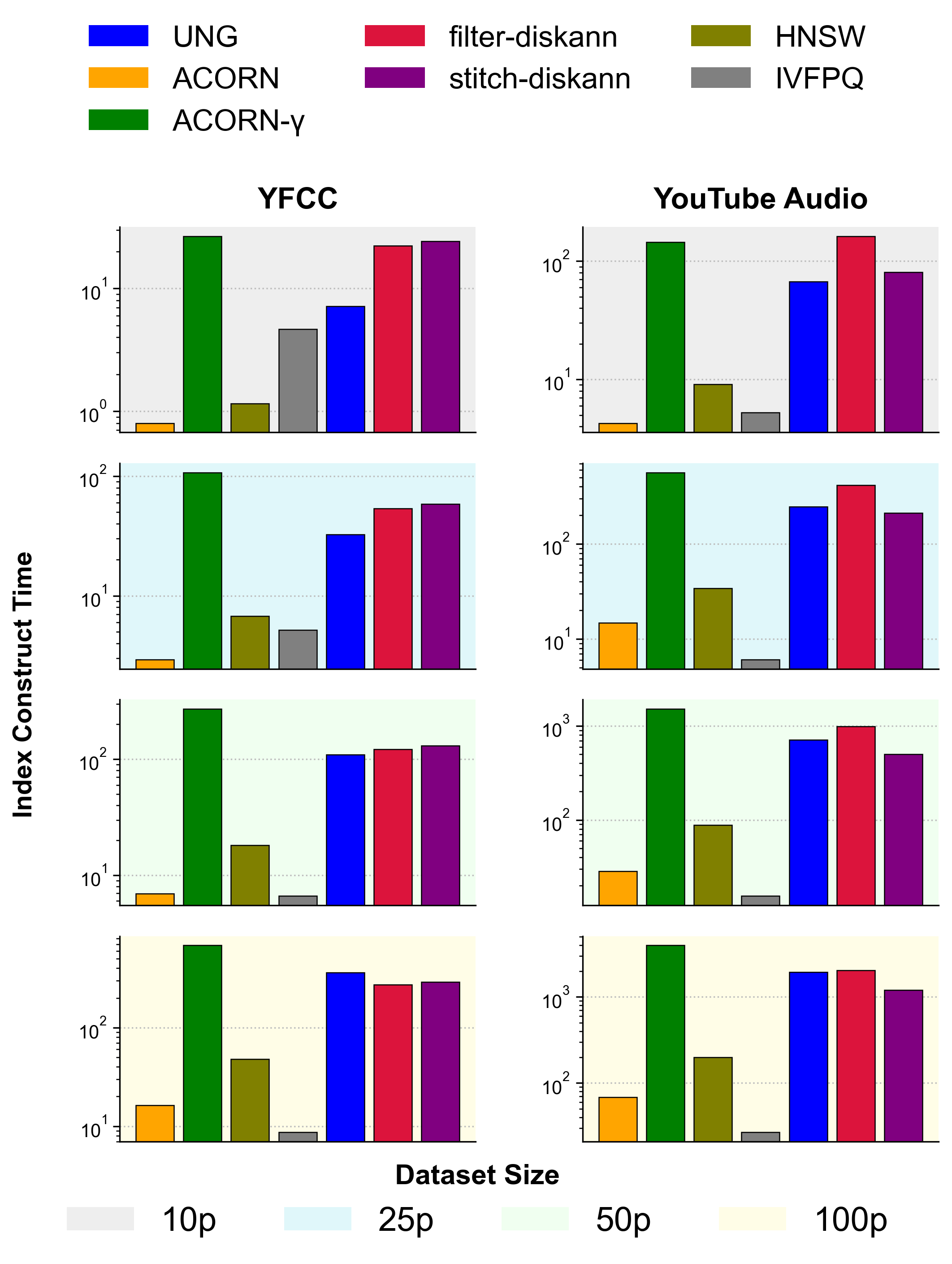}
    \vspace{-7mm}
    \caption{Effect of Varying Base Datasets Size on index construct time.}
    \label{fig: base index time}
    \vspace{-3mm}
\end{figure}

Figure~\ref{fig: base index time} illustrates the impact of dataset size scaling on indexing time across the YouTube-Audio and YFCC datasets. While the indexing time of most algorithms increases linearly as the dataset size grows, there are notable differences between the two datasets. Specifically, algorithms including \texttt{ACORN-1}, \texttt{HNSW}, and \texttt{IVFPQ} exhibit similar and relatively short indexing times, enabling efficient index construction—additionally, they do not require considering label distribution during the indexing process. In contrast, the remaining algorithms share comparable indexing times that are approximately one order of magnitude higher. Notably, these algorithms generally need to account for label distribution when building the index.

\section{Recommendation}

\textbf{Method Evaluation} We recommend evaluating methods primarily via QPS--Recall trade-offs and reporting Pareto-frontier points under consistent parameter sweeps. Assess stability across stratified query difficulties, including (i) label-length groups, (ii) selectivity percentiles, and (iii) varying top-$k$; note that varying the base dataset size typically has limited impact on search performance. Conduct experiments on representative datasets with diverse label distributions and cover comprehensive scenarios (\textit{containment}, \textit{equality}, \textit{overlap}). Where appropriate, include index build time and memory footprint to decouple retrieval latency from construction cost, and report reproducible best-parameter sets for each method. Because many real-world datasets exhibit very low selectivity, it is crucial to emphasize evaluation under low-selectivity conditions, where throughput and recall often diverge most across methods.

\noindent \textbf{Method Selection} 

\begin{itemize}[leftmargin=*]
    \item Prefer \texttt{UNG} as the default for strong QPS--recall in Containment and Equality scenarios; under very low selectivity or long labels, filter-then-search (e.g., \texttt{UNG}, \texttt{ACORN-1}, or even pre-filter Brute Force) keeps candidates small and throughput high; at high $k$, favor \texttt{UNG}/\texttt{ACORN} and avoid post-filter \texttt{IVFPQ}.
  \item Fixed-Length Equality is an easier, structured case where most methods reach high recall with proper tuning; \texttt{UNG} and \texttt{CAPS} offer top QPS at high recall.
  \item Prefer \texttt{ACORN-1} and \texttt{stitch-diskann} in Overlap scenario for balanced throughput and recall; \texttt{ACORN} and post-filter \texttt{HNSW} are robust for large label spaces or high $k$; avoid \texttt{UNG} as query becomes more complex causing QPS/recall degradation.
  \item At low $k$ (e.g., recall@1), post-filter methods and \texttt{UNG} are competitive with high QPS; at higher $k$ (e.g., recall@100), prefer \texttt{UNG}/\texttt{ACORN} for containment/equality and \texttt{DiskANN}/\texttt{ACORN} for overlap.
  \item If indexing time or memory is constrained, \texttt{HNSW}, \texttt{IVFPQ}, and \texttt{ACORN-1} build quickly, are more memory-efficient, and offer greater flexibility as base labels change. Methods that rely on predicate label distributions (e.g., \texttt{UNG}, \texttt{DiskANN} variants) may incur higher build costs.
\end{itemize}

\section{Conclusion}

\label{sec: conclusion}

In this paper, we conducted a comprehensive benchmark and in-depth analysis of filtered Approximate Nearest Neighbor Search (FANNS). Our work addressed three obstacles: the combinatorial explosion of algorithmic configurations, the insufficient understanding of detailed impact factors, and the widespread fragmentation of evaluation methodologies. To overcome these challenges, we introduced a novel evaluation framework. By categorizing algorithms and employing a unified parameter tuning strategy, our framework mitigates evaluation bias and enables a fair, parameter-aware comparison that highlights the trade-offs of different algorithms. Through extensive experiments, we provided an in-depth analysis of how performance is affected by diverse factors, revealing the practical strengths and operational boundaries of each approach under realistic conditions.

\bibliographystyle{ACM-Reference-Format}
\bibliography{sample}


\begin{thebibliography}{53}


\ifx \showCODEN    \undefined \def \showCODEN     #1{\unskip}     \fi
\ifx \showDOI      \undefined \def \showDOI       #1{#1}\fi
\ifx \showISBNx    \undefined \def \showISBNx     #1{\unskip}     \fi
\ifx \showISBNxiii \undefined \def \showISBNxiii  #1{\unskip}     \fi
\ifx \showISSN     \undefined \def \showISSN      #1{\unskip}     \fi
\ifx \showLCCN     \undefined \def \showLCCN      #1{\unskip}     \fi
\ifx \shownote     \undefined \def \shownote      #1{#1}          \fi
\ifx \showarticletitle \undefined \def \showarticletitle #1{#1}   \fi
\ifx \showURL      \undefined \def \showURL       {\relax}        \fi
\providecommand\bibfield[2]{#2}
\providecommand\bibinfo[2]{#2}
\providecommand\natexlab[1]{#1}
\providecommand\showeprint[2][]{arXiv:#2}

\bibitem[\protect\citeauthoryear{Abu-El-Haija, Kothari, Lee, Natsev, Toderici, Varadarajan, and Vijayanarasimhan}{Abu-El-Haija et~al\mbox{.}}{2016}]%
        {youtube}
\bibfield{author}{\bibinfo{person}{Sami Abu-El-Haija}, \bibinfo{person}{Nisarg Kothari}, \bibinfo{person}{Joonseok Lee}, \bibinfo{person}{Paul Natsev}, \bibinfo{person}{George Toderici}, \bibinfo{person}{Balakrishnan Varadarajan}, {and} \bibinfo{person}{Sudheendra Vijayanarasimhan}.} \bibinfo{year}{2016}\natexlab{}.
\newblock \bibinfo{title}{YouTube-8M: A Large-Scale Video Classification Benchmark}.
\newblock
\newblock
\showeprint[arxiv]{1609.08675}~[cs.CV]
\urldef\tempurl%
\url{https://arxiv.org/abs/1609.08675}
\showURL{%
\tempurl}


\bibitem[\protect\citeauthoryear{Boguñá, Krioukov, and Claffy}{Boguñá et~al\mbox{.}}{2008}]%
        {nsw2}
\bibfield{author}{\bibinfo{person}{Marián Boguñá}, \bibinfo{person}{Dmitri Krioukov}, {and} \bibinfo{person}{K.~C. Claffy}.} \bibinfo{year}{2008}\natexlab{}.
\newblock \showarticletitle{Navigability of complex networks}.
\newblock \bibinfo{journal}{\emph{Nature Physics}} \bibinfo{volume}{5}, \bibinfo{number}{1} (\bibinfo{date}{Nov.} \bibinfo{year}{2008}), \bibinfo{pages}{74–80}.
\newblock
\showISSN{1745-2481}
\urldef\tempurl%
\url{https://doi.org/10.1038/nphys1130}
\showDOI{\tempurl}


\bibitem[\protect\citeauthoryear{Cai, Shi, Chen, and Zheng}{Cai et~al\mbox{.}}{2024}]%
        {UNG}
\bibfield{author}{\bibinfo{person}{Yuzheng Cai}, \bibinfo{person}{Jiayang Shi}, \bibinfo{person}{Yizhuo Chen}, {and} \bibinfo{person}{Weiguo Zheng}.} \bibinfo{year}{2024}\natexlab{}.
\newblock \showarticletitle{Navigating Labels and Vectors: A Unified Approach to Filtered Approximate Nearest Neighbor Search}.
\newblock \bibinfo{journal}{\emph{Proc. ACM Manag. Data}} \bibinfo{volume}{2}, \bibinfo{number}{6}, Article \bibinfo{articleno}{246} (\bibinfo{date}{Dec.} \bibinfo{year}{2024}), \bibinfo{numpages}{27}~pages.
\newblock
\urldef\tempurl%
\url{https://doi.org/10.1145/3698822}
\showDOI{\tempurl}


\bibitem[\protect\citeauthoryear{Coleman, Segarra, Shrivastava, and Smola}{Coleman et~al\mbox{.}}{2021}]%
        {graph_reorder}
\bibfield{author}{\bibinfo{person}{Benjamin Coleman}, \bibinfo{person}{Santiago Segarra}, \bibinfo{person}{Anshumali Shrivastava}, {and} \bibinfo{person}{Alex Smola}.} \bibinfo{year}{2021}\natexlab{}.
\newblock \bibinfo{title}{Graph Reordering for Cache-Efficient Near Neighbor Search}.
\newblock
\newblock
\showeprint[arxiv]{2104.03221}~[cs.DS]
\urldef\tempurl%
\url{https://arxiv.org/abs/2104.03221}
\showURL{%
\tempurl}


\bibitem[\protect\citeauthoryear{Dong, Moses, and Li}{Dong et~al\mbox{.}}{2011}]%
        {kgraph}
\bibfield{author}{\bibinfo{person}{Wei Dong}, \bibinfo{person}{Charikar Moses}, {and} \bibinfo{person}{Kai Li}.} \bibinfo{year}{2011}\natexlab{}.
\newblock \showarticletitle{Efficient k-nearest neighbor graph construction for generic similarity measures}. In \bibinfo{booktitle}{\emph{Proceedings of the 20th International Conference on World Wide Web}} (Hyderabad, India) \emph{(\bibinfo{series}{WWW '11})}. \bibinfo{publisher}{Association for Computing Machinery}, \bibinfo{address}{New York, NY, USA}, \bibinfo{pages}{577–586}.
\newblock
\showISBNx{9781450306324}
\urldef\tempurl%
\url{https://doi.org/10.1145/1963405.1963487}
\showDOI{\tempurl}


\bibitem[\protect\citeauthoryear{Douze, Guzhva, Deng, Johnson, Szilvasy, Mazaré, Lomeli, Hosseini, and Jégou}{Douze et~al\mbox{.}}{2025}]%
        {faiss}
\bibfield{author}{\bibinfo{person}{Matthijs Douze}, \bibinfo{person}{Alexandr Guzhva}, \bibinfo{person}{Chengqi Deng}, \bibinfo{person}{Jeff Johnson}, \bibinfo{person}{Gergely Szilvasy}, \bibinfo{person}{Pierre-Emmanuel Mazaré}, \bibinfo{person}{Maria Lomeli}, \bibinfo{person}{Lucas Hosseini}, {and} \bibinfo{person}{Hervé Jégou}.} \bibinfo{year}{2025}\natexlab{}.
\newblock \bibinfo{title}{The Faiss library}.
\newblock
\newblock
\showeprint[arxiv]{2401.08281}~[cs.LG]
\urldef\tempurl%
\url{https://arxiv.org/abs/2401.08281}
\showURL{%
\tempurl}


\bibitem[\protect\citeauthoryear{Echihabi, Zoumpatianos, and Palpanas}{Echihabi et~al\mbox{.}}{2021}]%
        {ANNSsurvey1}
\bibfield{author}{\bibinfo{person}{Karima Echihabi}, \bibinfo{person}{Kostas Zoumpatianos}, {and} \bibinfo{person}{Themis Palpanas}.} \bibinfo{year}{2021}\natexlab{}.
\newblock \showarticletitle{New trends in high-d vector similarity search: al-driven, progressive, and distributed}.
\newblock \bibinfo{journal}{\emph{Proceedings of the VLDB Endowment}} \bibinfo{volume}{14}, \bibinfo{number}{12} (\bibinfo{year}{2021}), \bibinfo{pages}{3198--3201}.
\newblock


\bibitem[\protect\citeauthoryear{Elastic}{Elastic}{[n.d.]}]%
        {elastic2025elasticsearch}
\bibfield{author}{\bibinfo{person}{Elastic}.} \bibinfo{year}{[n.d.]}\natexlab{}.
\newblock \bibinfo{booktitle}{\emph{Elasticsearch}}.
\newblock
\urldef\tempurl%
\url{https://github.com/elastic/elasticsearch}
\showURL{%
\tempurl}


\bibitem[\protect\citeauthoryear{Fu, Xiang, Wang, and Cai}{Fu et~al\mbox{.}}{2025}]%
        {NSG}
\bibfield{author}{\bibinfo{person}{Cong Fu}, \bibinfo{person}{Chao Xiang}, \bibinfo{person}{Changxu Wang}, {and} \bibinfo{person}{Deng Cai}.} \bibinfo{year}{2025}\natexlab{}.
\newblock \bibinfo{title}{Fast Approximate Nearest Neighbor Search With The Navigating Spreading-out Graph}.
\newblock
\newblock
\showeprint[arxiv]{1707.00143}~[cs.LG]
\urldef\tempurl%
\url{https://arxiv.org/abs/1707.00143}
\showURL{%
\tempurl}


\bibitem[\protect\citeauthoryear{Gao, Gou, Xu, Yang, Long, and Wong}{Gao et~al\mbox{.}}{2024}]%
        {extended-rabitq}
\bibfield{author}{\bibinfo{person}{Jianyang Gao}, \bibinfo{person}{Yutong Gou}, \bibinfo{person}{Yuexuan Xu}, \bibinfo{person}{Yongyi Yang}, \bibinfo{person}{Cheng Long}, {and} \bibinfo{person}{Raymond Chi-Wing Wong}.} \bibinfo{year}{2024}\natexlab{}.
\newblock \bibinfo{title}{Practical and Asymptotically Optimal Quantization of High-Dimensional Vectors in Euclidean Space for Approximate Nearest Neighbor Search}.
\newblock
\newblock
\showeprint[arxiv]{2409.09913}~[cs.DB]
\urldef\tempurl%
\url{https://arxiv.org/abs/2409.09913}
\showURL{%
\tempurl}


\bibitem[\protect\citeauthoryear{Gao and Long}{Gao and Long}{2024}]%
        {rabitq}
\bibfield{author}{\bibinfo{person}{Jianyang Gao} {and} \bibinfo{person}{Cheng Long}.} \bibinfo{year}{2024}\natexlab{}.
\newblock \showarticletitle{RaBitQ: Quantizing High-Dimensional Vectors with a Theoretical Error Bound for Approximate Nearest Neighbor Search}.
\newblock \bibinfo{journal}{\emph{Proc. ACM Manag. Data}} \bibinfo{volume}{2}, \bibinfo{number}{3}, Article \bibinfo{articleno}{167} (\bibinfo{date}{May} \bibinfo{year}{2024}), \bibinfo{numpages}{27}~pages.
\newblock
\urldef\tempurl%
\url{https://doi.org/10.1145/3654970}
\showDOI{\tempurl}


\bibitem[\protect\citeauthoryear{Gao, Xiong, Gao, Jia, Pan, Bi, Dai, Sun, and Wang}{Gao et~al\mbox{.}}{2023}]%
        {RAGsurvey2023}
\bibfield{author}{\bibinfo{person}{Yunfan Gao}, \bibinfo{person}{Yun Xiong}, \bibinfo{person}{Xinyu Gao}, \bibinfo{person}{Kangxiang Jia}, \bibinfo{person}{Jinliu Pan}, \bibinfo{person}{Yuxi Bi}, \bibinfo{person}{Yi Dai}, \bibinfo{person}{Jiawei Sun}, {and} \bibinfo{person}{Haofen Wang}.} \bibinfo{year}{2023}\natexlab{}.
\newblock \showarticletitle{Retrieval-augmented generation for large language models: A survey}.
\newblock \bibinfo{journal}{\emph{arXiv preprint arXiv:2312.10997}} (\bibinfo{year}{2023}).
\newblock


\bibitem[\protect\citeauthoryear{Ge, He, Ke, and Sun}{Ge et~al\mbox{.}}{2014}]%
        {OPQ}
\bibfield{author}{\bibinfo{person}{Tiezheng Ge}, \bibinfo{person}{Kaiming He}, \bibinfo{person}{Qifa Ke}, {and} \bibinfo{person}{Jian Sun}.} \bibinfo{year}{2014}\natexlab{}.
\newblock \showarticletitle{Optimized Product Quantization}.
\newblock \bibinfo{journal}{\emph{IEEE Transactions on Pattern Analysis and Machine Intelligence}} \bibinfo{volume}{36}, \bibinfo{number}{4} (\bibinfo{year}{2014}), \bibinfo{pages}{744--755}.
\newblock
\urldef\tempurl%
\url{https://doi.org/10.1109/TPAMI.2013.240}
\showDOI{\tempurl}


\bibitem[\protect\citeauthoryear{Gollapudi, Karia, Sivashankar, Krishnaswamy, Begwani, Raz, Lin, Zhang, Mahapatro, Srinivasan, et~al\mbox{.}}{Gollapudi et~al\mbox{.}}{2023}]%
        {filterDiskANN}
\bibfield{author}{\bibinfo{person}{Siddharth Gollapudi}, \bibinfo{person}{Neel Karia}, \bibinfo{person}{Varun Sivashankar}, \bibinfo{person}{Ravishankar Krishnaswamy}, \bibinfo{person}{Nikit Begwani}, \bibinfo{person}{Swapnil Raz}, \bibinfo{person}{Yiyong Lin}, \bibinfo{person}{Yin Zhang}, \bibinfo{person}{Neelam Mahapatro}, \bibinfo{person}{Premkumar Srinivasan}, {et~al\mbox{.}}} \bibinfo{year}{2023}\natexlab{}.
\newblock \showarticletitle{Filtered-diskann: Graph algorithms for approximate nearest neighbor search with filters}. In \bibinfo{booktitle}{\emph{Proceedings of the ACM Web Conference 2023}}. \bibinfo{pages}{3406--3416}.
\newblock


\bibitem[\protect\citeauthoryear{Gupta}{Gupta}{2018}]%
        {capscode}
\bibfield{author}{\bibinfo{person}{Gaurav Gupta}.} \bibinfo{year}{2018}\natexlab{}.
\newblock \bibinfo{booktitle}{\emph{constrainedANN}}.
\newblock
\urldef\tempurl%
\url{https://github.com/gaurav16gupta/constrainedANN}
\showURL{%
\tempurl}


\bibitem[\protect\citeauthoryear{Gupta, Yi, Coleman, Luo, Lakshman, and Shrivastava}{Gupta et~al\mbox{.}}{2023}]%
        {CAPS}
\bibfield{author}{\bibinfo{person}{Gaurav Gupta}, \bibinfo{person}{Jonah Yi}, \bibinfo{person}{Benjamin Coleman}, \bibinfo{person}{Chen Luo}, \bibinfo{person}{Vihan Lakshman}, {and} \bibinfo{person}{Anshumali Shrivastava}.} \bibinfo{year}{2023}\natexlab{}.
\newblock \showarticletitle{CAPS: A Practical Partition Index for Filtered Similarity Search}.
\newblock \bibinfo{journal}{\emph{arXiv preprint arXiv:2308.15014}} (\bibinfo{year}{2023}).
\newblock


\bibitem[\protect\citeauthoryear{Huang, He, Gao, Deng, Acero, and Heck}{Huang et~al\mbox{.}}{2013}]%
        {dssm}
\bibfield{author}{\bibinfo{person}{Po-Sen Huang}, \bibinfo{person}{Xiaodong He}, \bibinfo{person}{Jianfeng Gao}, \bibinfo{person}{Li Deng}, \bibinfo{person}{Alex Acero}, {and} \bibinfo{person}{Larry Heck}.} \bibinfo{year}{2013}\natexlab{}.
\newblock \showarticletitle{Learning deep structured semantic models for web search using clickthrough data}. In \bibinfo{booktitle}{\emph{Proceedings of the 22nd ACM International Conference on Information \& Knowledge Management}} (San Francisco, California, USA) \emph{(\bibinfo{series}{CIKM '13})}. \bibinfo{publisher}{Association for Computing Machinery}, \bibinfo{address}{New York, NY, USA}, \bibinfo{pages}{2333–2338}.
\newblock
\showISBNx{9781450322638}
\urldef\tempurl%
\url{https://doi.org/10.1145/2505515.2505665}
\showDOI{\tempurl}


\bibitem[\protect\citeauthoryear{Jiang, Li, Zhu, De~Fine~Licht, He, Shi, Renggli, Zhang, Rekatsinas, Hoefler, and Alonso}{Jiang et~al\mbox{.}}{2023}]%
        {searchengine3}
\bibfield{author}{\bibinfo{person}{Wenqi Jiang}, \bibinfo{person}{Shigang Li}, \bibinfo{person}{Yu Zhu}, \bibinfo{person}{Johannes De~Fine~Licht}, \bibinfo{person}{Zhenhao He}, \bibinfo{person}{Runbin Shi}, \bibinfo{person}{Cedric Renggli}, \bibinfo{person}{Shuai Zhang}, \bibinfo{person}{Theodoros Rekatsinas}, \bibinfo{person}{Torsten Hoefler}, {and} \bibinfo{person}{Gustavo Alonso}.} \bibinfo{year}{2023}\natexlab{}.
\newblock \showarticletitle{Co-design Hardware and Algorithm for Vector Search}. In \bibinfo{booktitle}{\emph{Proceedings of the International Conference for High Performance Computing, Networking, Storage and Analysis}} (Denver, CO, USA) \emph{(\bibinfo{series}{SC '23})}. \bibinfo{publisher}{Association for Computing Machinery}, \bibinfo{address}{New York, NY, USA}, Article \bibinfo{articleno}{87}, \bibinfo{numpages}{15}~pages.
\newblock
\showISBNx{9798400701092}
\urldef\tempurl%
\url{https://doi.org/10.1145/3581784.3607045}
\showDOI{\tempurl}


\bibitem[\protect\citeauthoryear{Jégou, Douze, and Schmid}{Jégou et~al\mbox{.}}{2011a}]%
        {ivfpq}
\bibfield{author}{\bibinfo{person}{Herve Jégou}, \bibinfo{person}{Matthijs Douze}, {and} \bibinfo{person}{Cordelia Schmid}.} \bibinfo{year}{2011}\natexlab{a}.
\newblock \showarticletitle{Product Quantization for Nearest Neighbor Search}.
\newblock \bibinfo{journal}{\emph{IEEE Transactions on Pattern Analysis and Machine Intelligence}} \bibinfo{volume}{33}, \bibinfo{number}{1} (\bibinfo{year}{2011}), \bibinfo{pages}{117--128}.
\newblock
\urldef\tempurl%
\url{https://doi.org/10.1109/TPAMI.2010.57}
\showDOI{\tempurl}


\bibitem[\protect\citeauthoryear{Jégou, Douze, and Schmid}{Jégou et~al\mbox{.}}{2011b}]%
        {PQ}
\bibfield{author}{\bibinfo{person}{Herve Jégou}, \bibinfo{person}{Matthijs Douze}, {and} \bibinfo{person}{Cordelia Schmid}.} \bibinfo{year}{2011}\natexlab{b}.
\newblock \showarticletitle{Product Quantization for Nearest Neighbor Search}.
\newblock \bibinfo{journal}{\emph{IEEE Transactions on Pattern Analysis and Machine Intelligence}} \bibinfo{volume}{33}, \bibinfo{number}{1} (\bibinfo{year}{2011}), \bibinfo{pages}{117--128}.
\newblock
\urldef\tempurl%
\url{https://doi.org/10.1109/TPAMI.2010.57}
\showDOI{\tempurl}


\bibitem[\protect\citeauthoryear{{KGLab-HDU}}{{KGLab-HDU}}{2022}]%
        {NHQcode}
\bibfield{author}{\bibinfo{person}{{KGLab-HDU}}.} \bibinfo{year}{2022}\natexlab{}.
\newblock \bibinfo{booktitle}{\emph{{TKDE-under-review-Native-Hybrid-Queries-via-ANNS}}}.
\newblock
\urldef\tempurl%
\url{https://github.com/KGLab-HDU/TKDE-under-review-Native-Hybrid-Queries-via-ANNS}
\showURL{%
\tempurl}


\bibitem[\protect\citeauthoryear{Kleinberg}{Kleinberg}{2000}]%
        {nsw1}
\bibfield{author}{\bibinfo{person}{Jon Kleinberg}.} \bibinfo{year}{2000}\natexlab{}.
\newblock \showarticletitle{Kleinberg, J. Navigation in a small world. Nature 406, 845}.
\newblock \bibinfo{journal}{\emph{Nature}}  \bibinfo{volume}{406} (\bibinfo{date}{09} \bibinfo{year}{2000}), \bibinfo{pages}{845}.
\newblock
\urldef\tempurl%
\url{https://doi.org/10.1038/35022643}
\showDOI{\tempurl}


\bibitem[\protect\citeauthoryear{Lakshman, Teo, Chu, Nigam, Patni, Maknikar, and Vishwanathan}{Lakshman et~al\mbox{.}}{[n.d.]}]%
        {recommandation3}
\bibfield{author}{\bibinfo{person}{Vihan Lakshman}, \bibinfo{person}{ChoonHui Teo}, \bibinfo{person}{Xiaowen Chu}, \bibinfo{person}{Priyanka Nigam}, \bibinfo{person}{Abhinandan Patni}, \bibinfo{person}{Pooja Maknikar}, {and} \bibinfo{person}{SVN Vishwanathan}.} \bibinfo{year}{[n.d.]}\natexlab{}.
\newblock \showarticletitle{Embracing Structure in Data for Billion-Scale Semantic Product Search}.
\newblock  (\bibinfo{year}{[n.\,d.]}).
\newblock


\bibitem[\protect\citeauthoryear{Li, Liu, Wu, Xu, Zhao, Huang, Kang, Chen, Li, and Lee}{Li et~al\mbox{.}}{2019a}]%
        {mind}
\bibfield{author}{\bibinfo{person}{Chao Li}, \bibinfo{person}{Zhiyuan Liu}, \bibinfo{person}{Mengmeng Wu}, \bibinfo{person}{Yuchi Xu}, \bibinfo{person}{Huan Zhao}, \bibinfo{person}{Pipei Huang}, \bibinfo{person}{Guoliang Kang}, \bibinfo{person}{Qiwei Chen}, \bibinfo{person}{Wei Li}, {and} \bibinfo{person}{Dik~Lun Lee}.} \bibinfo{year}{2019}\natexlab{a}.
\newblock \showarticletitle{Multi-Interest Network with Dynamic Routing for Recommendation at Tmall}. In \bibinfo{booktitle}{\emph{Proceedings of the 28th ACM International Conference on Information and Knowledge Management}} (Beijing, China) \emph{(\bibinfo{series}{CIKM '19})}. \bibinfo{publisher}{Association for Computing Machinery}, \bibinfo{address}{New York, NY, USA}, \bibinfo{pages}{2615–2623}.
\newblock
\showISBNx{9781450369763}
\urldef\tempurl%
\url{https://doi.org/10.1145/3357384.3357814}
\showDOI{\tempurl}


\bibitem[\protect\citeauthoryear{Li, Lv, Jin, Lin, Yang, Zeng, Wu, and Ma}{Li et~al\mbox{.}}{2021}]%
        {recommandation2}
\bibfield{author}{\bibinfo{person}{Sen Li}, \bibinfo{person}{Fuyu Lv}, \bibinfo{person}{Taiwei Jin}, \bibinfo{person}{Guli Lin}, \bibinfo{person}{Keping Yang}, \bibinfo{person}{Xiaoyi Zeng}, \bibinfo{person}{Xiao-Ming Wu}, {and} \bibinfo{person}{Qianli Ma}.} \bibinfo{year}{2021}\natexlab{}.
\newblock \showarticletitle{Embedding-based Product Retrieval in Taobao Search}. In \bibinfo{booktitle}{\emph{Proceedings of the 27th ACM SIGKDD Conference on Knowledge Discovery \& Data Mining}}.
\newblock
\urldef\tempurl%
\url{https://doi.org/10.1145/3447548.3467101}
\showDOI{\tempurl}


\bibitem[\protect\citeauthoryear{Li, Zhang, Sun, Wang, Li, Zhang, and Lin}{Li et~al\mbox{.}}{2019b}]%
        {ANNSsurvey2}
\bibfield{author}{\bibinfo{person}{Wen Li}, \bibinfo{person}{Ying Zhang}, \bibinfo{person}{Yifang Sun}, \bibinfo{person}{Wei Wang}, \bibinfo{person}{Mingjie Li}, \bibinfo{person}{Wenjie Zhang}, {and} \bibinfo{person}{Xuemin Lin}.} \bibinfo{year}{2019}\natexlab{b}.
\newblock \showarticletitle{Approximate nearest neighbor search on high dimensional data—experiments, analyses, and improvement}.
\newblock \bibinfo{journal}{\emph{IEEE Transactions on Knowledge and Data Engineering}} \bibinfo{volume}{32}, \bibinfo{number}{8} (\bibinfo{year}{2019}), \bibinfo{pages}{1475--1488}.
\newblock


\bibitem[\protect\citeauthoryear{Liang, Zhang, Yao, Chen, Song, and Cheng}{Liang et~al\mbox{.}}{2025}]%
        {unify}
\bibfield{author}{\bibinfo{person}{Anqi Liang}, \bibinfo{person}{Pengcheng Zhang}, \bibinfo{person}{Bin Yao}, \bibinfo{person}{Zhongpu Chen}, \bibinfo{person}{Yitong Song}, {and} \bibinfo{person}{Guangxu Cheng}.} \bibinfo{year}{2025}\natexlab{}.
\newblock \bibinfo{title}{UNIFY: Unified Index for Range Filtered Approximate Nearest Neighbors Search}.
\newblock
\newblock
\showeprint[arxiv]{2412.02448}~[cs.DS]
\urldef\tempurl%
\url{https://arxiv.org/abs/2412.02448}
\showURL{%
\tempurl}


\bibitem[\protect\citeauthoryear{Magnani, Liu, Chaidaroon, Yadav, Reddy~Suram, Puthenputhussery, Chen, Xie, Kashi, Lee, and Liao}{Magnani et~al\mbox{.}}{2022}]%
        {ecommerce1}
\bibfield{author}{\bibinfo{person}{Alessandro Magnani}, \bibinfo{person}{Feng Liu}, \bibinfo{person}{Suthee Chaidaroon}, \bibinfo{person}{Sachin Yadav}, \bibinfo{person}{Praveen Reddy~Suram}, \bibinfo{person}{Ajit Puthenputhussery}, \bibinfo{person}{Sijie Chen}, \bibinfo{person}{Min Xie}, \bibinfo{person}{Anirudh Kashi}, \bibinfo{person}{Tony Lee}, {and} \bibinfo{person}{Ciya Liao}.} \bibinfo{year}{2022}\natexlab{}.
\newblock \showarticletitle{Semantic Retrieval at Walmart}. In \bibinfo{booktitle}{\emph{Proceedings of the 28th ACM SIGKDD Conference on Knowledge Discovery and Data Mining}} (Washington DC, USA) \emph{(\bibinfo{series}{KDD '22})}. \bibinfo{publisher}{Association for Computing Machinery}, \bibinfo{address}{New York, NY, USA}, \bibinfo{pages}{3495–3503}.
\newblock
\showISBNx{9781450393850}
\urldef\tempurl%
\url{https://doi.org/10.1145/3534678.3539164}
\showDOI{\tempurl}


\bibitem[\protect\citeauthoryear{Malkov and Yashunin}{Malkov and Yashunin}{2018}]%
        {HNSW}
\bibfield{author}{\bibinfo{person}{Yu~A Malkov} {and} \bibinfo{person}{Dmitry~A Yashunin}.} \bibinfo{year}{2018}\natexlab{}.
\newblock \showarticletitle{Efficient and robust approximate nearest neighbor search using hierarchical navigable small world graphs}.
\newblock \bibinfo{journal}{\emph{IEEE transactions on pattern analysis and machine intelligence}} \bibinfo{volume}{42}, \bibinfo{number}{4} (\bibinfo{year}{2018}), \bibinfo{pages}{824--836}.
\newblock


\bibitem[\protect\citeauthoryear{{Malteos}}{{Malteos}}{2022}]%
        {Arxiv}
\bibfield{author}{\bibinfo{person}{{Malteos}}.} \bibinfo{year}{2022}\natexlab{}.
\newblock \bibinfo{title}{Aspect Paper Embeddings}.
\newblock \bibinfo{howpublished}{\url{https://huggingface.co/datasets/malteos/aspect-paper-embeddings}}.
\newblock


\bibitem[\protect\citeauthoryear{Ni, Xu, Wang, Li, Yao, Xiao, and Zhang}{Ni et~al\mbox{.}}{2023}]%
        {diskann++}
\bibfield{author}{\bibinfo{person}{Jiongkang Ni}, \bibinfo{person}{Xiaoliang Xu}, \bibinfo{person}{Yuxiang Wang}, \bibinfo{person}{Can Li}, \bibinfo{person}{Jiajie Yao}, \bibinfo{person}{Shihai Xiao}, {and} \bibinfo{person}{Xuecang Zhang}.} \bibinfo{year}{2023}\natexlab{}.
\newblock \bibinfo{title}{DiskANN++: Efficient Page-based Search over Isomorphic Mapped Graph Index using Query-sensitivity Entry Vertex}.
\newblock
\newblock
\showeprint[arxiv]{2310.00402}~[cs.IR]
\urldef\tempurl%
\url{https://arxiv.org/abs/2310.00402}
\showURL{%
\tempurl}


\bibitem[\protect\citeauthoryear{Nigam, Song, Mohan, Lakshman, Ding, Shingavi, Teo, Gu, and Yin}{Nigam et~al\mbox{.}}{2019}]%
        {recommandation1}
\bibfield{author}{\bibinfo{person}{Priyanka Nigam}, \bibinfo{person}{Yiwei Song}, \bibinfo{person}{Vijai Mohan}, \bibinfo{person}{Vihan Lakshman}, \bibinfo{person}{Weitian~(Allen) Ding}, \bibinfo{person}{Ankit Shingavi}, \bibinfo{person}{Choon~Hui Teo}, \bibinfo{person}{Hao Gu}, {and} \bibinfo{person}{Bing Yin}.} \bibinfo{year}{2019}\natexlab{}.
\newblock \showarticletitle{Semantic Product Search}. In \bibinfo{booktitle}{\emph{Proceedings of the 25th ACM SIGKDD International Conference on Knowledge Discovery \& Data Mining}}.
\newblock
\urldef\tempurl%
\url{https://doi.org/10.1145/3292500.3330759}
\showDOI{\tempurl}


\bibitem[\protect\citeauthoryear{Niu, Li, Li, Xiao, Sun, Wang, Deng, and Chen}{Niu et~al\mbox{.}}{2020}]%
        {ecommerce2}
\bibfield{author}{\bibinfo{person}{Xichuan Niu}, \bibinfo{person}{Bofang Li}, \bibinfo{person}{Chenliang Li}, \bibinfo{person}{Rong Xiao}, \bibinfo{person}{Haochuan Sun}, \bibinfo{person}{Honggang Wang}, \bibinfo{person}{Hongbo Deng}, {and} \bibinfo{person}{Zhenzhong Chen}.} \bibinfo{year}{2020}\natexlab{}.
\newblock \showarticletitle{Gated Heterogeneous Graph Representation Learning for Shop Search in E-commerce}. In \bibinfo{booktitle}{\emph{Proceedings of the 29th ACM International Conference on Information \& Knowledge Management}} (Virtual Event, Ireland) \emph{(\bibinfo{series}{CIKM '20})}. \bibinfo{publisher}{Association for Computing Machinery}, \bibinfo{address}{New York, NY, USA}, \bibinfo{pages}{2165–2168}.
\newblock
\showISBNx{9781450368599}
\urldef\tempurl%
\url{https://doi.org/10.1145/3340531.3412087}
\showDOI{\tempurl}


\bibitem[\protect\citeauthoryear{{OpenMP Architecture Review Board}}{{OpenMP Architecture Review Board}}{2008}]%
        {openmp08}
\bibfield{author}{\bibinfo{person}{{OpenMP Architecture Review Board}}.} \bibinfo{year}{2008}\natexlab{}.
\newblock \bibinfo{title}{{OpenMP} Application Program Interface Version 3.0}.
\newblock
\newblock
\urldef\tempurl%
\url{http://www.openmp.org/mp-documents/spec30.pdf}
\showURL{%
\tempurl}


\bibitem[\protect\citeauthoryear{Ostendorff, Blume, Ruas, Gipp, and Rehm}{Ostendorff et~al\mbox{.}}{2022}]%
        {arxivembedding}
\bibfield{author}{\bibinfo{person}{Malte Ostendorff}, \bibinfo{person}{Till Blume}, \bibinfo{person}{Terry Ruas}, \bibinfo{person}{Bela Gipp}, {and} \bibinfo{person}{Georg Rehm}.} \bibinfo{year}{2022}\natexlab{}.
\newblock \showarticletitle{Specialized document embeddings for aspect-based similarity of research papers}. In \bibinfo{booktitle}{\emph{Proceedings of the 22nd ACM/IEEE Joint Conference on Digital Libraries}}. \bibinfo{pages}{1--12}.
\newblock


\bibitem[\protect\citeauthoryear{Patel}{Patel}{2024}]%
        {ACORN-codes}
\bibfield{author}{\bibinfo{person}{Liana Patel}.} \bibinfo{year}{2024}\natexlab{}.
\newblock \bibinfo{title}{ACORN}.
\newblock \bibinfo{howpublished}{\url{https://github.com/stanford-futuredata/ACORN}}.
\newblock


\bibitem[\protect\citeauthoryear{Patel, Kraft, Guestrin, and Zaharia}{Patel et~al\mbox{.}}{2024}]%
        {acorn}
\bibfield{author}{\bibinfo{person}{Liana Patel}, \bibinfo{person}{Peter Kraft}, \bibinfo{person}{Carlos Guestrin}, {and} \bibinfo{person}{Matei Zaharia}.} \bibinfo{year}{2024}\natexlab{}.
\newblock \showarticletitle{ACORN: Performant and Predicate-Agnostic Search Over Vector Embeddings and Structured Data}.
\newblock \bibinfo{journal}{\emph{Proc. ACM Manag. Data}} \bibinfo{volume}{2}, \bibinfo{number}{3}, Article \bibinfo{articleno}{120} (\bibinfo{date}{May} \bibinfo{year}{2024}), \bibinfo{numpages}{27}~pages.
\newblock
\urldef\tempurl%
\url{https://doi.org/10.1145/3654923}
\showDOI{\tempurl}


\bibitem[\protect\citeauthoryear{Radford, Kim, Hallacy, Ramesh, Goh, Agarwal, Sastry, Askell, Mishkin, Clark, Krueger, and Sutskever}{Radford et~al\mbox{.}}{2021}]%
        {CLIP}
\bibfield{author}{\bibinfo{person}{Alec Radford}, \bibinfo{person}{Jong~Wook Kim}, \bibinfo{person}{Chris Hallacy}, \bibinfo{person}{Aditya Ramesh}, \bibinfo{person}{Gabriel Goh}, \bibinfo{person}{Sandhini Agarwal}, \bibinfo{person}{Girish Sastry}, \bibinfo{person}{Amanda Askell}, \bibinfo{person}{Pamela Mishkin}, \bibinfo{person}{Jack Clark}, \bibinfo{person}{Gretchen Krueger}, {and} \bibinfo{person}{Ilya Sutskever}.} \bibinfo{year}{2021}\natexlab{}.
\newblock \showarticletitle{Learning Transferable Visual Models From Natural Language Supervision}.
\newblock \bibinfo{journal}{\emph{CoRR}}  \bibinfo{volume}{abs/2103.00020} (\bibinfo{year}{2021}).
\newblock
\showeprint[arXiv]{2103.00020}
\urldef\tempurl%
\url{https://arxiv.org/abs/2103.00020}
\showURL{%
\tempurl}


\bibitem[\protect\citeauthoryear{Rekabsaz, Lesota, Schedl, Brassey, and Eickhoff}{Rekabsaz et~al\mbox{.}}{2021}]%
        {tripclick}
\bibfield{author}{\bibinfo{person}{Navid Rekabsaz}, \bibinfo{person}{Oleg Lesota}, \bibinfo{person}{Markus Schedl}, \bibinfo{person}{Jon Brassey}, {and} \bibinfo{person}{Carsten Eickhoff}.} \bibinfo{year}{2021}\natexlab{}.
\newblock \showarticletitle{Tripclick: The Log Files of a Large Health Web Search Engine}. In \bibinfo{booktitle}{\emph{Proceedings of the 44th International ACM SIGIR Conference on Research and Development in Information Retrieval}}. \bibinfo{pages}{2507--2513}.
\newblock


\bibitem[\protect\citeauthoryear{Schuhmann, Vencu, Beaumont, Kaczmarczyk, Mullis, Katta, Coombes, Jitsev, and Komatsuzaki}{Schuhmann et~al\mbox{.}}{2021}]%
        {LAION}
\bibfield{author}{\bibinfo{person}{Christoph Schuhmann}, \bibinfo{person}{Richard Vencu}, \bibinfo{person}{Romain Beaumont}, \bibinfo{person}{Robert Kaczmarczyk}, \bibinfo{person}{Clayton Mullis}, \bibinfo{person}{Aarush Katta}, \bibinfo{person}{Theo Coombes}, \bibinfo{person}{Jenia Jitsev}, {and} \bibinfo{person}{Aran Komatsuzaki}.} \bibinfo{year}{2021}\natexlab{}.
\newblock \showarticletitle{LAION-400M: Open Dataset of Clip-filtered 400 Million Image-text Pairs}.
\newblock \bibinfo{journal}{\emph{arXiv preprint arXiv:2111.02114}} (\bibinfo{year}{2021}).
\newblock


\bibitem[\protect\citeauthoryear{Simhadri, Aumüller, Ingber, Douze, Williams, Manohar, Baranchuk, Liberty, Liu, Landrum, Karjikar, Dhulipala, Chen, Chen, Ma, Zhang, Cai, Shi, Chen, Zheng, Wan, Yin, and Huang}{Simhadri et~al\mbox{.}}{2024}]%
        {bigann}
\bibfield{author}{\bibinfo{person}{Harsha~Vardhan Simhadri}, \bibinfo{person}{Martin Aumüller}, \bibinfo{person}{Amir Ingber}, \bibinfo{person}{Matthijs Douze}, \bibinfo{person}{George Williams}, \bibinfo{person}{Magdalen~Dobson Manohar}, \bibinfo{person}{Dmitry Baranchuk}, \bibinfo{person}{Edo Liberty}, \bibinfo{person}{Frank Liu}, \bibinfo{person}{Ben Landrum}, \bibinfo{person}{Mazin Karjikar}, \bibinfo{person}{Laxman Dhulipala}, \bibinfo{person}{Meng Chen}, \bibinfo{person}{Yue Chen}, \bibinfo{person}{Rui Ma}, \bibinfo{person}{Kai Zhang}, \bibinfo{person}{Yuzheng Cai}, \bibinfo{person}{Jiayang Shi}, \bibinfo{person}{Yizhuo Chen}, \bibinfo{person}{Weiguo Zheng}, \bibinfo{person}{Zihao Wan}, \bibinfo{person}{Jie Yin}, {and} \bibinfo{person}{Ben Huang}.} \bibinfo{year}{2024}\natexlab{}.
\newblock \bibinfo{title}{Results of the Big ANN: NeurIPS'23 competition}.
\newblock
\newblock
\showeprint[arxiv]{2409.17424}~[cs.IR]
\urldef\tempurl%
\url{https://arxiv.org/abs/2409.17424}
\showURL{%
\tempurl}


\bibitem[\protect\citeauthoryear{Simhadri, Krishnaswamy, Srinivasa, Subramanya, Antonijevic, Pryce, Kaczynski, Williams, Gollapudi, Sivashankar, Karia, Singh, Jaiswal, Mahapatro, Adams, Tower, and Patel}{Simhadri et~al\mbox{.}}{[n.d.]}]%
        {diskann-github}
\bibfield{author}{\bibinfo{person}{Harsha~Vardhan Simhadri}, \bibinfo{person}{Ravishankar Krishnaswamy}, \bibinfo{person}{Gopal Srinivasa}, \bibinfo{person}{Suhas~Jayaram Subramanya}, \bibinfo{person}{Andrija Antonijevic}, \bibinfo{person}{Dax Pryce}, \bibinfo{person}{David Kaczynski}, \bibinfo{person}{Shane Williams}, \bibinfo{person}{Siddarth Gollapudi}, \bibinfo{person}{Varun Sivashankar}, \bibinfo{person}{Neel Karia}, \bibinfo{person}{Aditi Singh}, \bibinfo{person}{Shikhar Jaiswal}, \bibinfo{person}{Neelam Mahapatro}, \bibinfo{person}{Philip Adams}, \bibinfo{person}{Bryan Tower}, {and} \bibinfo{person}{Yash Patel}.} \bibinfo{year}{[n.d.]}\natexlab{}.
\newblock
\newblock


\bibitem[\protect\citeauthoryear{Singh and Dwivedi}{Singh and Dwivedi}{2015}]%
        {searchengine4}
\bibfield{author}{\bibinfo{person}{Jitendra~Nath Singh} {and} \bibinfo{person}{Sanjay~K. Dwivedi}.} \bibinfo{year}{2015}\natexlab{}.
\newblock \showarticletitle{Performance Evaluation of Search Engines Using Enhanced Vector Space Model}.
\newblock \bibinfo{journal}{\emph{Journal of Computer Science}} \bibinfo{volume}{11}, \bibinfo{number}{4} (\bibinfo{date}{Jul} \bibinfo{year}{2015}), \bibinfo{pages}{692--698}.
\newblock
\urldef\tempurl%
\url{https://doi.org/10.3844/jcssp.2015.692.698}
\showDOI{\tempurl}


\bibitem[\protect\citeauthoryear{Subramanya, Devvrit, Simhadri, Krishnawamy, and Kadekodi}{Subramanya et~al\mbox{.}}{2019b}]%
        {Vamana}
\bibfield{author}{\bibinfo{person}{SuhasJayaram Subramanya}, \bibinfo{person}{Fnu Devvrit}, \bibinfo{person}{HarshaVardhan Simhadri}, \bibinfo{person}{Ravishankar Krishnawamy}, {and} \bibinfo{person}{Rohan Kadekodi}.} \bibinfo{year}{2019}\natexlab{b}.
\newblock \showarticletitle{DiskANN: Fast Accurate Billion-point Nearest Neighbor Search on a Single Node}.
\newblock \bibinfo{journal}{\emph{Neural Information Processing Systems,Neural Information Processing Systems}} (\bibinfo{date}{Nov} \bibinfo{year}{2019}).
\newblock


\bibitem[\protect\citeauthoryear{Subramanya, Devvrit, Kadekodi, Krishaswamy, and Simhadri}{Subramanya et~al\mbox{.}}{2019a}]%
        {diskann}
\bibfield{author}{\bibinfo{person}{Suhas~Jayaram Subramanya}, \bibinfo{person}{Devvrit}, \bibinfo{person}{Rohan Kadekodi}, \bibinfo{person}{Ravishankar Krishaswamy}, {and} \bibinfo{person}{Harsha~Vardhan Simhadri}.} \bibinfo{year}{2019}\natexlab{a}.
\newblock \bibinfo{booktitle}{\emph{DiskANN: fast accurate billion-point nearest neighbor search on a single node}}.
\newblock \bibinfo{publisher}{Curran Associates Inc.}, \bibinfo{address}{Red Hook, NY, USA}.
\newblock


\bibitem[\protect\citeauthoryear{Thomee, Shamma, Friedland, Elizalde, Ni, Poland, Borth, and Li}{Thomee et~al\mbox{.}}{2016}]%
        {YFCC}
\bibfield{author}{\bibinfo{person}{Bart Thomee}, \bibinfo{person}{David~A. Shamma}, \bibinfo{person}{Gerald Friedland}, \bibinfo{person}{Benjamin Elizalde}, \bibinfo{person}{Karl Ni}, \bibinfo{person}{Douglas Poland}, \bibinfo{person}{Damian Borth}, {and} \bibinfo{person}{Li-Jia Li}.} \bibinfo{year}{2016}\natexlab{}.
\newblock \showarticletitle{YFCC100M: the new data in multimedia research}.
\newblock \bibinfo{journal}{\emph{Commun. ACM}} \bibinfo{volume}{59}, \bibinfo{number}{2} (\bibinfo{date}{Jan.} \bibinfo{year}{2016}), \bibinfo{pages}{64–73}.
\newblock
\showISSN{0001-0782}
\urldef\tempurl%
\url{https://doi.org/10.1145/2812802}
\showDOI{\tempurl}


\bibitem[\protect\citeauthoryear{Wang, Lv, Xu, Wang, Yue, and Ni}{Wang et~al\mbox{.}}{2024}]%
        {NHQ}
\bibfield{author}{\bibinfo{person}{Mengzhao Wang}, \bibinfo{person}{Lingwei Lv}, \bibinfo{person}{Xiaoliang Xu}, \bibinfo{person}{Yuxiang Wang}, \bibinfo{person}{Qiang Yue}, {and} \bibinfo{person}{Jiongkang Ni}.} \bibinfo{year}{2024}\natexlab{}.
\newblock \showarticletitle{An efficient and robust framework for approximate nearest neighbor search with attribute constraint}.
\newblock \bibinfo{journal}{\emph{Advances in Neural Information Processing Systems}}  \bibinfo{volume}{36} (\bibinfo{year}{2024}).
\newblock


\bibitem[\protect\citeauthoryear{Wang, Xu, Yue, and Wang}{Wang et~al\mbox{.}}{2021}]%
        {ANNSsurvey3}
\bibfield{author}{\bibinfo{person}{Mengzhao Wang}, \bibinfo{person}{Xiaoliang Xu}, \bibinfo{person}{Qiang Yue}, {and} \bibinfo{person}{Yuxiang Wang}.} \bibinfo{year}{2021}\natexlab{}.
\newblock \showarticletitle{A comprehensive survey and experimental comparison of graph-based approximate nearest neighbor search}.
\newblock \bibinfo{journal}{\emph{arXiv preprint arXiv:2101.12631}} (\bibinfo{year}{2021}).
\newblock


\bibitem[\protect\citeauthoryear{Wu, He, Qiao, Fu, Liu, and Yu}{Wu et~al\mbox{.}}{2022}]%
        {HQANN}
\bibfield{author}{\bibinfo{person}{Wei Wu}, \bibinfo{person}{Junlin He}, \bibinfo{person}{Yu Qiao}, \bibinfo{person}{Guoheng Fu}, \bibinfo{person}{Li Liu}, {and} \bibinfo{person}{Jin Yu}.} \bibinfo{year}{2022}\natexlab{}.
\newblock \showarticletitle{HQANN: Efficient and robust similarity search for hybrid queries with structured and unstructured constraints}. In \bibinfo{booktitle}{\emph{Proceedings of the 31st ACM International Conference on Information \& Knowledge Management}}. \bibinfo{pages}{4580--4584}.
\newblock


\bibitem[\protect\citeauthoryear{Yang, Hu, Peng, Li, Li, Wang, and Liu}{Yang et~al\mbox{.}}{2024}]%
        {vdtuner}
\bibfield{author}{\bibinfo{person}{Tiannuo Yang}, \bibinfo{person}{Wen Hu}, \bibinfo{person}{Wangqi Peng}, \bibinfo{person}{Yusen Li}, \bibinfo{person}{Jianguo Li}, \bibinfo{person}{Gang Wang}, {and} \bibinfo{person}{Xiaoguang Liu}.} \bibinfo{year}{2024}\natexlab{}.
\newblock \showarticletitle{VDTuner: Automated Performance Tuning for Vector Data Management Systems}. In \bibinfo{booktitle}{\emph{2024 IEEE 40th International Conference on Data Engineering (ICDE)}}. \bibinfo{pages}{4357--4369}.
\newblock
\urldef\tempurl%
\url{https://doi.org/10.1109/ICDE60146.2024.00332}
\showDOI{\tempurl}


\bibitem[\protect\citeauthoryear{Zhao, Zhang, Yu, Wang, Geng, Fu, Yang, Zhang, and Cui}{Zhao et~al\mbox{.}}{2024}]%
        {RAGsurvey2024}
\bibfield{author}{\bibinfo{person}{Penghao Zhao}, \bibinfo{person}{Hailin Zhang}, \bibinfo{person}{Qinhan Yu}, \bibinfo{person}{Zhengren Wang}, \bibinfo{person}{Yunteng Geng}, \bibinfo{person}{Fangcheng Fu}, \bibinfo{person}{Ling Yang}, \bibinfo{person}{Wentao Zhang}, {and} \bibinfo{person}{Bin Cui}.} \bibinfo{year}{2024}\natexlab{}.
\newblock \showarticletitle{Retrieval-Augmented Generation for AI-Generated Content: A Survey}.
\newblock \bibinfo{journal}{\emph{arXiv preprint arXiv:2402.19473}} (\bibinfo{year}{2024}).
\newblock


\bibitem[\protect\citeauthoryear{Zuo and Deng}{Zuo and Deng}{2023}]%
        {arkgraph}
\bibfield{author}{\bibinfo{person}{Chaoji Zuo} {and} \bibinfo{person}{Dong Deng}.} \bibinfo{year}{2023}\natexlab{}.
\newblock \showarticletitle{ARKGraph: All-Range Approximate K-Nearest-Neighbor Graph}.
\newblock \bibinfo{journal}{\emph{Proc. VLDB Endow.}} \bibinfo{volume}{16}, \bibinfo{number}{10} (\bibinfo{date}{June} \bibinfo{year}{2023}), \bibinfo{pages}{2645–2658}.
\newblock
\showISSN{2150-8097}
\urldef\tempurl%
\url{https://doi.org/10.14778/3603581.3603601}
\showDOI{\tempurl}


\bibitem[\protect\citeauthoryear{Zuo, Qiao, Zhou, Li, and Deng}{Zuo et~al\mbox{.}}{2024}]%
        {serf}
\bibfield{author}{\bibinfo{person}{Chaoji Zuo}, \bibinfo{person}{Miao Qiao}, \bibinfo{person}{Wenchao Zhou}, \bibinfo{person}{Feifei Li}, {and} \bibinfo{person}{Dong Deng}.} \bibinfo{year}{2024}\natexlab{}.
\newblock \showarticletitle{SeRF: Segment Graph for Range-Filtering Approximate Nearest Neighbor Search}.
\newblock \bibinfo{journal}{\emph{Proc. ACM Manag. Data}} \bibinfo{volume}{2}, \bibinfo{number}{1}, Article \bibinfo{articleno}{69} (\bibinfo{date}{March} \bibinfo{year}{2024}), \bibinfo{numpages}{26}~pages.
\newblock
\urldef\tempurl%
\url{https://doi.org/10.1145/3639324}
\showDOI{\tempurl}


\end{thebibliography}

\end{document}